\documentclass[10pt,aps,amsmath,amssymb, prd,nofootinbib, twocolumn,preprintnumbers,superscriptaddress,authoryear,longbibliography]{revtex4-2}

\usepackage{graphicx}
\usepackage{amsmath}
\usepackage[caption=false]{subfig}
\usepackage{siunitx}
\usepackage{placeins}
\usepackage{color}
\usepackage{standalone}
\usepackage{dcolumn}
\usepackage{tensor}
\usepackage{bm}
\usepackage{microtype}
\usepackage{etoolbox}
\usepackage{amssymb}
\usepackage{mathrsfs}
\usepackage{accents}
\usepackage[normalem]{ulem}
\usepackage{soul} 
\usepackage[dvipsnames]{xcolor}
\usepackage[colorlinks,urlcolor=Blue,citecolor=Blue,linkcolor=Blue,pdfusetitle]{hyperref}
\usepackage[all]{hypcap}
\usepackage[inline]{enumitem}
\usepackage[utf8]{inputenc}
\usepackage{xspace}
\usepackage[printonlyused, nolist]{acronym}
\usepackage{lineno}
\usepackage{bbding}
\usepackage{pifont}
\usepackage{wasysym}
\usepackage{xspace}
\usepackage{multirow}

\usepackage{tablefootnote}
\usepackage{booktabs}
\usepackage{longtable}
\usepackage{color, colortbl}
\definecolor{Color1}{rgb}{0.9453,0.9453,0.9687}
\definecolor{Color2}{rgb}{0.8476,0.8945,0.9375}
\definecolor{Color3}{rgb}{0.7461,0.8164,0.8945}
\definecolor{Color4}{rgb}{0.6484,0.7383,0.8554}
\definecolor{Color5}{rgb}{0.5,0.6484,0.8047}

\usepackage{orcidlink}

\definecolor{electricviolet}{RGB}{143,0,255}

\definecolor{darkorange}{RGB}{255, 140, 0} 

\begin{document}

\title{Fast and accurate parameter estimation of high-redshift sources\\ with the Einstein Telescope}

\author{Filippo Santoliquido\,\orcidlink{0000-0003-3752-1400}}
\email{filippo.santoliquido@gssi.it}
\affiliation{Gran Sasso Science Institute (GSSI), I-67100 L'Aquila, Italy}
\affiliation{INFN, Laboratori Nazionali del Gran Sasso, I-67100 Assergi, Italy}

\author{Jacopo Tissino\,\orcidlink{0000-0003-2483-6710}}
\affiliation{Gran Sasso Science Institute (GSSI), I-67100 L'Aquila, Italy}
\affiliation{INFN, Laboratori Nazionali del Gran Sasso, I-67100 Assergi, Italy}

\author{Ulyana Dupletsa\,\orcidlink{0000-0003-2766-247X}}
\affiliation{Gran Sasso Science Institute (GSSI), I-67100 L'Aquila, Italy}
\affiliation{INFN, Laboratori Nazionali del Gran Sasso, I-67100 Assergi, Italy}
\affiliation{Institute of High Energy Physics - Austrian Academy of Sciences, 1010 Vienna, Austria}

\author{Marica Branchesi\,\orcidlink{0000-0003-1643-0526}}
\affiliation{Gran Sasso Science Institute (GSSI), I-67100 L'Aquila, Italy}
\affiliation{INFN, Laboratori Nazionali del Gran Sasso, I-67100 Assergi, Italy}

\author{Jan Harms\,\orcidlink{0000-0002-7332-9806}}
\affiliation{Gran Sasso Science Institute (GSSI), I-67100 L'Aquila, Italy}
\affiliation{INFN, Laboratori Nazionali del Gran Sasso, I-67100 Assergi, Italy}

\author{Manuel Arca Sedda\,\orcidlink{0000-0002-3987-0519}}
\affiliation{Gran Sasso Science Institute (GSSI), I-67100 L'Aquila, Italy}
\affiliation{INFN, Laboratori Nazionali del Gran Sasso, I-67100 Assergi, Italy}

\author{Maximilian Dax\,\orcidlink{0000-0001-8798-0627}}
\affiliation{ETH Zurich, CH-8092 Zurich, Switzerland}
\affiliation{ELLIS Institute Tübingen, D-72076 Tübingen, Germany}
\affiliation{Max Planck Institute for Intelligent Systems, D-72076 Tübingen, Germany}

\author{Annalena Kofler\,\orcidlink{0009-0008-5938-6215}}
\affiliation{Max Planck Institute for Intelligent Systems, D-72076 Tübingen, Germany}
\affiliation{Max Planck Institute for Gravitational Physics (Albert Einstein Institute), D-14476 Potsdam, Germany}

\author{Stephen R. Green\,\orcidlink{0000-0002-6987-6313}}
\affiliation{School of Mathematical Sciences, University of Nottingham, University Park, Nottingham, NG7 2RD, United Kingdom}

\author{Nihar Gupte\,\orcidlink{0000-0002-7287-5151}}
\affiliation{Max Planck Institute for Gravitational Physics (Albert Einstein Institute), D-14476 Potsdam, Germany}
\affiliation{Department of Physics, University of Maryland, College Park, MD 20742, USA}

\author{Isobel M. Romero-Shaw\,\orcidlink{0000-0002-4181-8090}} 
\affiliation{Department of Applied Mathematics and Theoretical Physics, Cambridge CB3 0WA, United Kingdom}
\affiliation{Kavli Institute for Cosmology Cambridge, Madingley Road Cambridge CB3 0HA, United Kingdom}
\affiliation{H.H. Wills Physics Laboratory, Tyndall Avenue, Bristol BS8 1TL, United Kingdom}

\author{Emanuele Berti\,\orcidlink{0000-0003-0751-5130}} 
\affiliation{William H. Miller III Department of Physics and Astronomy, Johns Hopkins University, Baltimore, Maryland 21218, USA}

\begin{abstract}

The Einstein Telescope (ET), along with other third-generation gravitational wave (GW) detectors, will be a key instrument for detecting GWs in the coming decades. However, analyzing the data and estimating source parameters will be challenging, especially given the large number of expected detections--on the order of $10^5$ per year--which makes current methods based on stochastic sampling impractical. In this work, we use {\sc Dingo-IS} to perform neural posterior estimation (NPE) of high-redshift events detectable with ET in its triangular configuration. NPE is a likelihood-free inference technique that leverages normalizing flows to approximate posterior distributions. After training, inference is fast, requiring only a few minutes per source, and accurate, as corrected through importance sampling and validated against standard Bayesian inference methods. To confirm previous findings on the ability to estimate parameters for high-redshift sources with ET, we compare NPE results with predictions from the Fisher information matrix (FIM) approximation. We find that NPE correctly recovers the eight degenerate sky modes induced by the triangular detector geometry, missed by the FIM analysis, resulting in an underestimation of sky localization uncertainties for most sources. FIM also overestimates the uncertainty in luminosity distance by a factor of $\sim 3$ on average when the injected luminosity distance is $d^{\mathrm{inj}}_{\mathrm{L}} > 10^5~$Mpc, further confirming that ET will be particularly well suited for studying the early Universe.


\end{abstract}

\maketitle
\section{Introduction}

The Einstein Telescope (ET) will be a leading instrument for detecting gravitational waves (GWs) in the next decades \citep{Punturo:2010zz, ET:2019dnz, Branchesi:2023mws,Abac:2025saz}, working alongside other third-generation (3G) GW observatories such as Cosmic Explorer \citep{Reitze:2019iox,Evans:2023euw,Gupta:2023lga} and LISA \citep{LISA:2024hlh}.

ET will be able to detect GWs with unprecedented sensitivity, but this also brings great challenges to data analysis and parameter estimation \citep[PE;][]{Couvares:2021ajn}. Some key issues include longer signals due to the wider frequency range, a large number of detections \citep[of order $10^5$ events per year;][]{Baibhav:2019gxm,2024PhRvD.110h3040B,ET:2019dnz,Branchesi:2023mws}, and overlapping signals \citep{2021PhRvD.104d4003S}. To address these challenges, new methods and technologies will be needed.

In the ET sensitivity band, 
a binary neutron star (BNS) signal can last from several minutes to hours. During this time, the noise characteristics of the detector may change \citep{2021PhRvD.104f3034Z,2021PhRvD.103l4061E,2022arXiv220212762K}, leading to non-stationary noise. Furthermore, transient noise signals, known as glitches, can overlap the signal in the data stream \citep{2015PhRvD..91h4034L,2015CQGra..32m5012C}, as also occurs in current GW detectors. Adding to the challenge, with the advent of 3G detectors, manually removing glitches that coincide with signals will no longer be practical, highlighting the need for the development of new methods to systematically and automatically handle non-Gaussian noise \citep{Narola:2024qdh,Legin:2024gid,Raymond:2024xzj,2025PhRvD.111b4019X,Sun:2023vlq}.

Not only can glitches overlap with signals, but signals can also overlap with each other \citep{2023PhRvL.130q1402L}. This presents another potential issue for data analysis in the ET era \citep{Wu:2022pyg,Janquart:2022fzz}. Parameter estimation becomes biased when the merger times of two overlapping signals are close, and accuracy worsens when their signal-to-noise ratios (SNRs) are similar \citep{2021PhRvD.104d4003S,2022PhRvD.105j4016P,2021PhRvD.104h4039R,2021PhRvD.104d4010H,2021MNRAS.507.5069A,2024CQGra..41e5011W,Papalini:2025exy}. 

The parameters $\theta$ describing a GW signal are estimated from data $d$ using Bayes' theorem \citep{vanderSluys:2007st,vanderSluys:2009bf,Thrane:2018qnx,Christensen:2022bxb}:  
\begin{equation}  
\label{eq:bayes}
    p(\theta|d) = \frac{\mathcal{L}(d|\theta)\pi(\theta)}{\mathcal{Z}(d)},  
\end{equation}  
where $\mathcal{L}(d|\theta)$ is the likelihood function, $\pi(\theta)$ is the prior distribution and $\mathcal{Z}(d) = \int \mathrm{d}\theta \mathcal{L}(d|\theta)\pi(\theta)$ is the evidence. The posterior distribution $p(\theta|d)$ is typically represented by a set of samples obtained with stochastic sampling methods \citep{Ashton:2018jfp}.

Several methods have been developed to accelerate stochastic sampling, including relative binning \citep[also called heterodyning;][]{2010arXiv1007.4820C,Veitch:2014wba,2021PhRvD.104l3030L}, multibanding \citep[or adaptive frequency resolution;][]{Vinciguerra:2017ngf,Aubin:2020goo,2021PhRvD.104d4062M,Allene:2025saz}, and reduced-order quadrature \citep{2015PhRvL.114g1104C, 2021PhRvL.127h1102S,2023PhRvD.107h4037T}. All of these aim to simplify likelihood evaluations and, thus accelerate convergence. These techniques have already been tested and shown to be effective for 3G detectors \citep{Baral:2025geo}, though under certain simplifying assumptions \citep{Smith:2021bqc}, such as ignoring Earth's rotation for long-duration signals \citep{2024PhRvD.110h4085N,2024arXiv241202651H}.

Recent years have seen significant advances in machine-learning-based analysis methods, resulting in substantial speedup in PE with precision comparable to traditional methods \citep{Cuoco:2020ogp,2020PhRvL.124d1102C, 2022NatPh..18..112G,2023arXiv231115585Z,Bhardwaj:2023xph,2024arXiv241215046C, 2024arXiv241203454H,2024arXiv241021076P,2024arXiv240107406S}.  

One such method is simulation-based inference \citep[or likelihood-free inference;][]{DBLP:journals/corr/abs-1905-07488,Cranmer_2020}, which bypasses the need for likelihood computation by leveraging the ability to generate data through simulations. The core idea in the GW context is to approximate the posterior distribution $p(\theta|d)$ in Eq.~\ref{eq:bayes} using normalizing flows \citep{Rezende:2015ocs,Papamakarios:2019fms,Kobyzev_2021}. These are invertible transformations parametrized by neural networks trained to transform a multivariate Gaussian base distribution into the target posterior distribution. This approach is also known as neural posterior estimation \citep[NPE;][]{Papamakarios:2016ctj}. Several previous studies have applied NPE or related methods in the context of GWs \citep{Chatterjee:2019gqr,Shen:2019vep,2020arXiv201012931D,2020PhRvD.102j4057G,Green:2020dnx,Krastev:2020skk,2021PhRvL.127x1103D,2023PhRvL.130q1403D,2023mla..confE..34W,2023PhRvD.108d2004B,2024arXiv240302443K,Mould:2025dts,Papalini:2025exy}. 

Normalizing flows can be used to accelerate stochastic sampling by improving the efficiency of sample proposals. For instance, they are integrated into nested sampling with {\sc{nessai}} \citep{2021PhRvD.103j3006W,2023MLS&T...4c5011W}, significantly reducing the number of required likelihood evaluations \citep{2020MNRAS.493.3132S}. Similarly, normalizing flows can enhance Markov Chain Monte Carlo (MCMC) methods, as implemented in {\sc{jim}} \citep{2023ApJ...958..129W,2024PhRvD.110h3033W}, which relies on {\sc{flowMC}} \citep{2023JOSS....8.5021W} and {\sc{ripple}} to speed up waveform generation 
and enable automatic
differentiation with {\sc{jax}} \citep{2024PhRvD.110f4028E,jax2018github}.

Additionally, NPE can accelerate population inference \citep{2024PhRvD.109f4056L}. Extracting population properties from a large set of detected events $\{d\}$ using hierarchical Bayesian analysis can be more computationally demanding than estimating parameters for individual events. Resolving this issue with new methods is crucial and will be necessary before the beginning of the 3G detector era \citep{2019PhRvD.100d3030T,gwpop_pipe,Mastrogiovanni:2023zbw,gwkokab2024github,Colloms:2025hib,Mancarella:2025uat,Mould:2025dts}.

The methods outlined above were developed primarily for current GW detectors. However, they hold significant promise for 3G detectors as well \citep{2024arXiv241202651H}.

In this decade, it is crucial to shape the design of 3G detectors to maximize their scientific output \citep{Branchesi:2023mws}. To guide decisions, large simulated populations of compact binary coalescences have been analyzed using the Fisher information matrix (FIM) approximation \citep{Dupletsa:2022scg,2023A&C....4500759D,2022ApJS..263....2I,Gupta:2023lga,Borhanian:2022czq,Begnoni:2025oyd}. However, it is now necessary to develop new methods that can perform full PE for sources detectable by 3G detectors.

This study has two main objectives. First, we apply and test NPE to provide full PE for sources detectable with the ET in its triangular configuration. We validate NPE by comparing its output with results obtained with standard inference libraries, such as {\sc Bilby} \citep{Ashton:2018jfp,Romero-Shaw:2020owr}. Second, we validate previous results on ET performance in estimating source parameters by comparing results from NPE and FIM, by running a large-scale injection campaign. While previous studies have examined uncertainties in sky localization and other parameters using standard Bayesian inference and FIM for second-generation detectors, like LIGO, Virgo \citep{Grover:2013sha,Rodriguez:2013mla,Dupletsa:2024gfl}, and LISA \citep{Porter:2015eha}, we are not aware of any that have done this thoroughly for ET.

We will focus on sources that merge at moderate and high redshift (i.e., 
$1 \lesssim z\lesssim 45$) for several reasons. First, these signals are short in the ET frequency band, lasting less than a few seconds, making them more accessible for training NPE. Second, high-redshift signals have low SNRs, making reliable PE using the FIM challenging \citep{2008PhRvD..77d2001V}.  Third, 
detecting and characterizing high-redshift sources is central to the scientific goals of 3G GW detectors \citep{Abac:2025saz}, highlighting the need for methods that can provide fast and accurate PE.

Several physical processes lead to the formation of high-redshift sources in our Universe. For example, Population~III (Pop.~III) stars, which form from metal-free gas at $z > 20$ \citep{Haiman1996,Tegmark1997,abel2002,yoshida2006,klessen2023}, are believed to have been the first stars to form and are thought therefore to produce the earliest stellar-origin black holes \citep{kinugawa2016,Hartwig2016,belczynski2017,tanikawa2022b,costa2023,Nandal2023,liu2023arXiv,Boyuan2024,tanikawa2024,Mestichelli:2024djn}. As a result, Pop.~III binary black holes (BBHs) have been extensively studied in recent years as sources of high-redshift GWs \citep{kinugawa2020,liubromm2020GW190521,liubromm2020,tanikawa2022,tanikawa2021b,wang2022,Plunkett:2025mjr}. In particular, \cite{santoliquido2023} found that over 20\% of detectable Pop.~III BBHs can occur at $z > 8$. Additionally, primordial black holes are expected to form at high redshift, covering a wide range of masses and merger rates \citep{DeLuca:2020qqa,DeLuca:2022iix,Franciolini:2022tfm,Ng:2021sqn,Ng:2022agi,Ng:2020qpk}.

This manuscript is organized as follows: Sec.~\ref{sec:methods} describes the methodology, including detector and waveform parametrization (Sec.~\ref{sec:detectors}), chosen priors (Sec.~\ref{sec:prior}), NPE (Sec.~\ref{sec:dingo}), nested sampling with {\sc Bilby} (Sec.~\ref{sec:bilby}), and the FIM approximation (Sec.~\ref{sec:fisher}). Results for a single source and population analysis are presented in Secs.~\ref{sec:single} and \ref{sec:pop}, respectively. The degeneracies in sky localization (Sec.~\ref{sec:skyloc}) and geocentric time (Sec.~\ref{sec:geocentric-time}) introduced by the triangular shape of ET are discussed first, followed by an analysis of the impact of higher-order modes and precessing spins on our results (Sec.~\ref{sec:hom}). In Sec.~\ref{sec:astro}, we estimate the parameters of injected sources based on an astrophysical population of BBH mergers formed from Pop.~III stars, while implications of our work on previous findings are discussed in Sec.~\ref{sec:impli}. Section~\ref{sec:energy} compares the estimated energy required for NPE and stochastic sampling to analyze a population of sources. We discuss caveats of our methods in Sec.~\ref{sec:caveats} and conclude in Sec.~\ref{sec:conclusions}.

\section{Methods}
\label{sec:methods}

\subsection{Detector and waveforms}
\label{sec:detectors}

\begin{figure}
    \centering
    \includegraphics[width=0.9\linewidth]{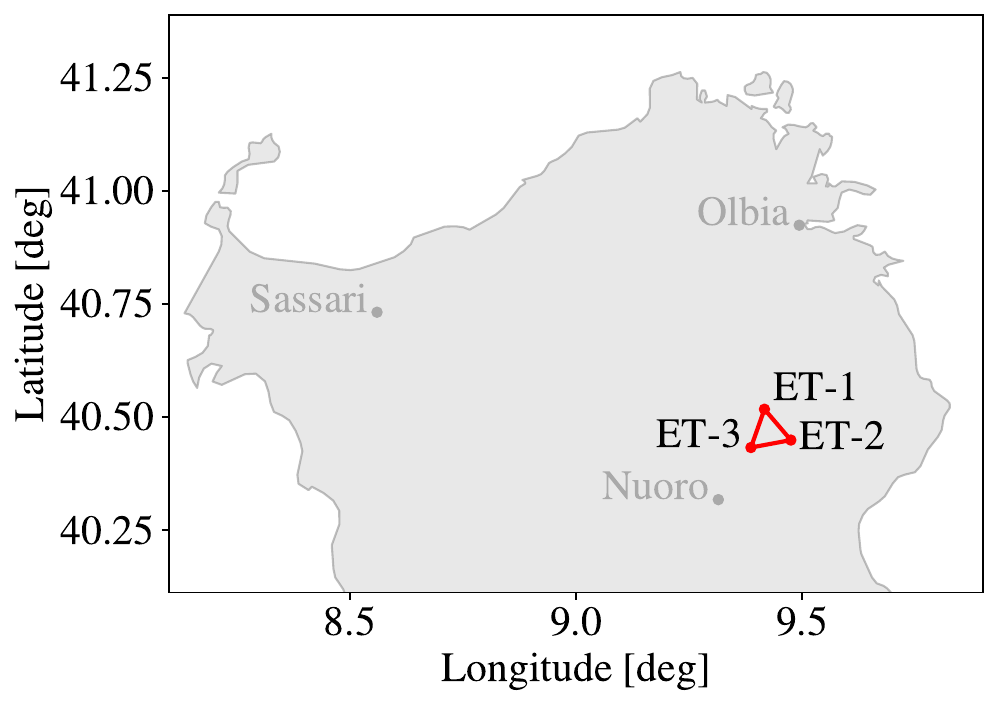}
    \caption{Locations and orientations of the three independent detectors (ET-1, ET-2, and ET-3) constituting ET-$\Delta$. See Sec.~\ref{sec:detectors} for further details.}
    \label{fig:ETmap}
\end{figure}

\begin{figure}
    \centering
    \includegraphics[width=0.9\linewidth]{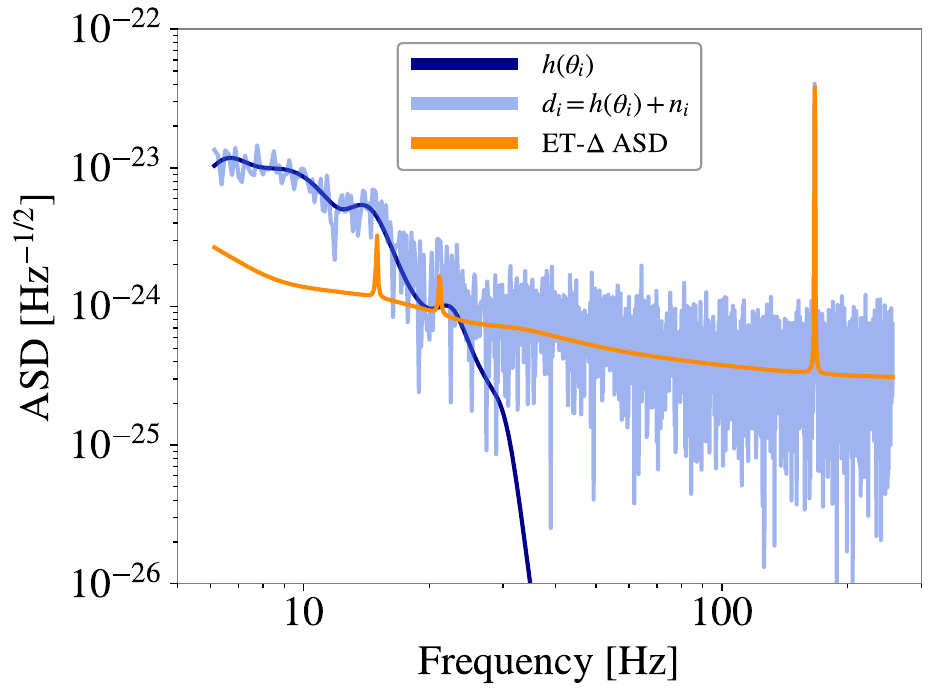}
    \caption{The injected signal $h(\theta_i)$ of \texttt{Event 1} projected on ET-1 in dark blue, the corresponding data $d_i$ after adding noise $n_i$ in light blue, and the adopted amplitude spectral density (ASD) in orange. See Sec~\ref{sec:detectors} and Sec.~\ref{sec:single} for further details.}
    \label{fig:data}
\end{figure}

We consider a single triangular ET detector, hereafter referred to as ET-$\Delta$, with an arm length of 10 km, located in Sardinia, which is a candidate site \citep{Branchesi:2023mws}. Figure~\ref{fig:ETmap} shows the locations of the three independent detectors that make up ET-$\Delta$, hereafter referred to as ET-1, ET-2, and ET-3, respectively. The triangular configuration is not the only proposed design for ET. Another option is the 2L configuration \citep{Branchesi:2023mws}, which consists of two L-shaped detectors with 15 km arms located in Europe. 

For our analysis, we use the HFLF cryogenic sensitivity curve \citep{psd}, which includes both a high-frequency instrument and a cryogenic low-frequency instrument. The estimated noise amplitude spectral density (ASD) is shown in Fig.~\ref{fig:data}. We also consider a frequency range between $f_{\mathrm{min}} = 6$ Hz and $f_{\mathrm{max}} = 256$ Hz. We use eight-second time segments, resulting in frequency bins of size $\mathrm{d}f = 0.125$ Hz.

The merger frequency, estimated as the frequency at which the inspiral phase ends $f_{\mathrm{ISCO}}$, is given by Eq.~14.108 in \cite{2018gwv..book.....M},
\begin{equation}
\label{eq:freq}
    f_{\mathrm{ISCO}} = 4.7~\mathrm{kHz}~ \frac{M_\odot}{M_{\mathrm{d}}} < 53~\mathrm{Hz},
\end{equation}
where $M_{\mathrm{d}} = 90$ M$_\odot$ is the total mass in the detector frame of the least massive binary considered in this study. 
$f_{\mathrm{ISCO}}$ is more than a factor of $4$ lower than $f_{\mathrm{max}}$, ensuring that the frequency band is wide enough to accommodate all the sources we analyze.

In a first approximation, the signal duration $\tau$ scales with the lower cutoff frequency $f_{\mathrm{min}}$ as given by Eq.~4.21 of \cite{Maggiore:2007ulw}:
\begin{equation}
\label{eq:duration}
    \tau \simeq 2.18\,\mathrm{s} \left ( \frac{1.21\,\mathrm{M}_\odot}{
    (1+z)\mathcal{M}_\mathrm{s}} \right)^{5/3} \left( \frac{100\,\mathrm{Hz}}{f_{\mathrm{min}}}\right)^{8/3},
\end{equation}
where $\mathcal{M}_\mathrm{s}$ is the source-frame chirp mass. By setting the frequency limit to $f_{\mathrm{min}}=6$ Hz, we ensure that the duration of all processed signals remains at most eight seconds. However, this choice may slightly underestimate ET's parameter estimation capabilities, whose sensitivity is expected to extend down to $\sim 2~\mathrm{Hz}$ (see Sec.~\ref{sec:caveats} for further discussion).

We define spin-aligned quasi-circular waveforms with parameters $\theta$ = $\{ \mathcal{M}_\mathrm{d}$, $q$, $d_{\mathrm{L}}$, $\mathrm{dec}$, $\mathrm{ra}$, $\theta_{\mathrm{jn}}$, $\psi$, $t_\mathrm{geocent}$, $\chi_1$, $\chi_2$, $\phi_c \}$, where $\mathcal{M}_\mathrm{d}$, $q$, and $d_{\mathrm{L}}$ denote the detector-frame chirp mass, mass ratio, and
luminosity distance, respectively; ra and dec denote the right ascension and declination
of the source's sky position,  $\theta_{\mathrm{jn}}$ is the angle between the the line of sight and the total angular momentum $\vec{J}$, $\psi$ is the polarization angle, $t_\mathrm{geocent}$ is the coalescence time at the center of the Earth, and $\phi_c$ is the phase at coalescence. The waveforms also depend on the aligned spin parameters $\chi_1$ and $\chi_2$, defined as $\chi_i = a_i\cos \theta_i$, where $a_i = |\chi_i|$ is the spin magnitude and $\theta_i$ is the angle with respect to the orbital angular momentum, defined as: 
\begin{equation}
\label{eq:aspins}
    \theta_i =
    \begin{cases} 
    0, & \text{if } \chi_i > 0, \\
    \pi, & \text{if } \chi_i < 0\,. \\
    \end{cases}
    \end{equation}
We use frequency-domain waveforms, fixing the reference frequency to \( f_{\mathrm{ref}} = 20 \) Hz \citep{Varma:2021csh}, with the waveform approximant {\sc{IMRPhenomXPHM}} \citep{PhysRevD.103.104056}, which models subdominant modes \([(\ell, |m|) = (2, 1), (3, 3), (3, 2), (4, 4)]\) in addition to the dominant mode \([(\ell, |m|) = (2, 2)]\). A detailed discussion on higher-order modes can be found in Sec.~\ref{sec:hom}.
 

\subsection{Priors}
\label{sec:prior}

\begin{table}
    \centering
    \begin{tabular}{lcr}
    \toprule
        Parameter & Units & Prior \\
        \hline
        $\mathcal{M}_{\mathrm{d}}$ & M$_\odot$ & Eq.~\ref{eq:mchirp} \\
        $q$ &  & Eq.~\ref{eq:massratio} \\
        $d_{\mathrm{L}}$ & Mpc & Eq.~\ref{eq:dl} \\
        ra & rad & $\mathcal{U}(0, 2\pi)$ \\
        $\cos$ dec &  & $\mathcal{U}(-1, 1)$ \\
        $\sin \theta_{\mathrm{jn}}$ &  & $\mathcal{U}(0, 1)$ \\
        $\phi_c$ & rad & $\mathcal{U}(0, 2\pi)$ \\
        $\psi$ & rad & $\mathcal{U}(0, \pi)$ \\
        $t_{\mathrm{geocent}}$ & s & $\mathcal{U}(-0.1, +0.1)$\\
        $\chi_i$ &  & Eq.~\ref{eq:spins}  \\

    \bottomrule    
    \end{tabular}
    \caption{GW signal parameters and their adopted priors. $\mathcal{U}(a, b)$ 
    represents a uniform 
    distribution between $a$ and $b$. See
Sec.~\ref{sec:prior} for details.}
    \label{tab:prior}
\end{table}

The prior distributions $\pi(\theta)$ adopted in Eq.~\ref{eq:bayes} are reported in Table~\ref{tab:prior}. In particular, for the detector-frame chirp mass and mass ratio, we choose a prior uniform in primary and secondary masses \citep{GWFish_Priors_Tutorial}, therefore:
\begin{equation}
    \label{eq:mchirp}
    \pi(\mathcal{M}_{\mathrm{d}}) \propto \mathcal{M}_{\mathrm{d}},
\end{equation}
where $\mathcal{M}_{\mathrm{d}} \in [40,~1100]$ M$_\odot$; and:
\begin{equation}
    \label{eq:massratio}
    \pi(q) \propto \frac{(1+q)^{2/5}}{q^{6/5}},
\end{equation}
where $q \in [0.125,~1]$. For the luminosity distance, we choose a prior uniform in
comoving volume and source-frame time,
\begin{equation}
\label{eq:dl}
    \pi(d_{\mathrm{L}}) \propto \frac{\mathrm{d} V_c}{\mathrm{d}d_{\mathrm{L}}}\frac{1}{1+z},
\end{equation}
where $d_{\mathrm{L}} \in [5 \times 10^3,~ 5 \times 10^5]$ Mpc. For the aligned spin component $\chi_i$, we form a joint prior as follows:
\begin{equation}
\label{eq:spins}
    \pi(\chi_i) = \int_{0}^{0.9}\mathrm{d}a_i  \int_{-1}^{1} \mathrm{d}\cos\theta_i \pi(a_i) \pi(\cos\theta_i) \delta(\chi_i - a_i\cos\theta_i),
\end{equation}
where $\pi(a_i) = \mathcal{U}(0,0.9)$ and $\cos\theta_i = \mathcal{U}(-1,1)$ \citep{lange2018rapidaccurateparameterinference,Ashton:2018jfp}.

By convention, we set the Earth orientation and the positions and orientations of the three interferometers to match those at the reference time \( t_{\mathrm{ref}} = 1126259462.391 \) s, which corresponds to the GPS trigger time of GW150914 \citep{Romero-Shaw:2020owr}. The merger time (\( t_{\mathrm{geocent}} \)) of all events is sampled randomly around this reference time. Fixing $t_{\mathrm{ref}}$ for such short signals is not a loss of generality, since any difference in time can be recovered.

In Appendix~\ref{app:priors}, we provide additional details on the prior and present a figure illustrating the distribution of the injected parameters.


\subsection{{\sc{Bilby}}}
\label{sec:bilby}

We compare the results obtained with NPE against {\sc Bilby} \citep{Romero-Shaw:2020owr}, which is a Python library specifically designed for Bayesian inference. Assuming the noise is stationary and Gaussian, the likelihood in Eq.~\ref{eq:bayes} follows \citep{Finn:1992wt}:
\begin{equation}
\label{eq:likelihood}
    \mathcal{L}(d|\theta) \propto \exp \left(  -\frac{1}{2} (d - h(\theta) | d - h(\theta)) \right),
\end{equation}
where $h(\theta)$ represents the gravitational wave signal as a function of frequency $f$ for a given set of parameters $\theta$, projected onto the antenna pattern of the detector (see Sec.~\ref{sec:skyloc} for details). Eq.~\ref{eq:likelihood} involves the noise-weighted inner product:
\begin{equation}
\label{eq:inner}
    (a|b) = 4~\mathrm{Re}~\int_{f_{\mathrm{min}}}^{f_{\mathrm{max}}} \mathrm{d}f \frac{a^*(f)b(f)}{S(f)},
\end{equation}
where $\mathrm{Re}$ denotes real part and $^*$ complex conjugate. Moreover, $\sqrt{S(f)}$ is the noise ASD, as described in Sec.~\ref{sec:detectors}, which takes into account that GW detectors are not equally sensitive at all frequencies.

To sample from Eq.~\ref{eq:bayes} with the likelihood defined in Eq.~\ref{eq:likelihood}, we use the nested sampling algorithm {\sc{nessai}} \citep{nessai,2021PhRvD.103j3006W,2023MLS&T...4c5011W}, which offers fast convergence by being accelerated with normalizing flows. 

Nested sampling methods produce posterior samples as a result of calculating the evidence $\mathcal{Z}(d)$ \citep{Skilling:2004pqw,Skilling:2006gxv,sivia2006data,Buchner:2021kpm,Ashton:2022grj}. Initially, a set of live points is drawn from the prior distribution. During each iteration, the live point with the lowest likelihood is replaced by a new sample. The algorithm ends once the remaining prior volume contributes less than a predefined fraction of the evidence \citep[e.g., \texttt{dlogz < 0.1};][]{Higson_2018,Speagle:2019ivv}. 


\subsection{{\sc{Dingo-IS}}}
\label{sec:dingo}

\begin{figure}
    \centering
    \includegraphics[width=0.9\linewidth]{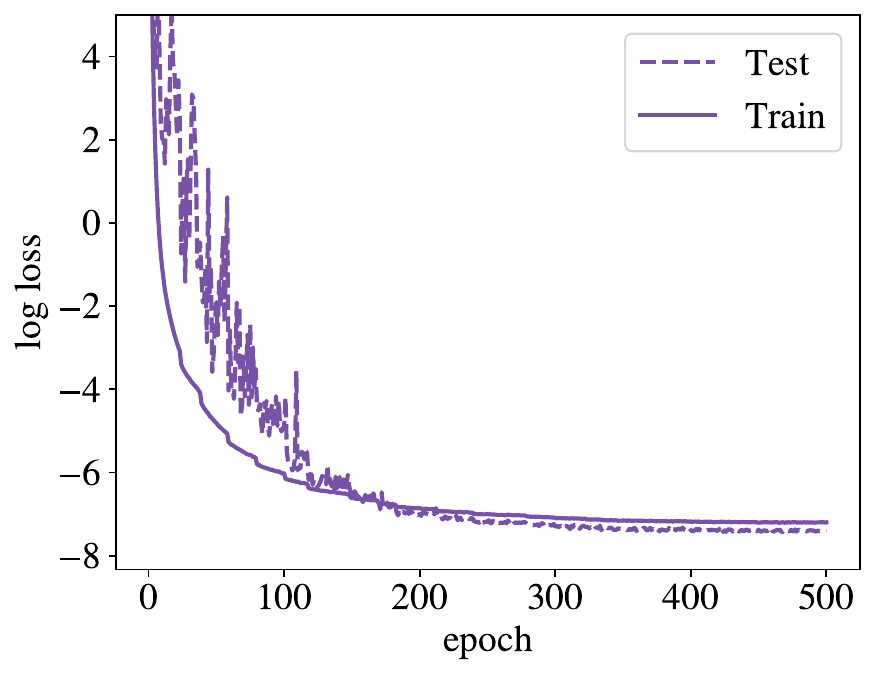}
    \caption{The $\log$ loss as a function of training epochs, shown as a solid line for the training set and a dashed line for the test set. See Sec.~\ref{sec:dingo} for details.}
    \label{fig:loss}
\end{figure}

To perform NPE, we use {\sc Dingo}, which estimates the conditional density $q(\theta|d)$ to approximate the true posterior $p(\theta|d)$. {\sc Dingo} is trained using simulated labeled strain data sets $(\theta, d)$, where $\theta$ is drawn from the prior $\pi(\theta)$ and $d$ is sampled from the likelihood $\mathcal{L}(d|\theta)$. This means that each strain data set consists of a waveform with additive stationary Gaussian noise, $d_i = h(\theta_i) + n_i$, where the noise realization is generated during training.

The {\sc Dingo} network architecture is the one presented in \cite{2021PhRvL.127x1103D}, with hyperparameters kept unchanged. In short, {\sc Dingo} consists of two neural networks: an embedding network and a normalizing flow. The embedding network compresses the input strain data to a set of 128 features, while the normalizing flows generate the Bayesian posterior conditioned on these features. Together, the two networks have more than $3 \times 10^8$ learnable parameters.

To train {\sc Dingo}, we generate 5 million waveforms with parameters extracted from the priors listed in Table~\ref{tab:prior} and the detector setup described in Sec.~\ref{sec:detectors}. To speed up training and inference, we simplify the process by avoiding the iterative group-equivariant NPE (GNPE) algorithm \citep[][see also Sec.~\ref{sec:hom}]{Dax:2021myb}  and fine-tuning for varying noise ASDs, since it is fixed in our case. We also analytically sample the phase parameter $\phi_c$ after training rather than trying to learn it. This is similar to phase marginalization \citep{Veitch:2014wba,Thrane:2018qnx} but does not require any approximations \citep{2023PhRvL.130q1403D}. 

During training, we set aside 2\% of the dataset for testing. To reduce overfitting, we use a scheduler that lowers the learning rate when the loss function stops improving  
after ten epochs \citep{NEURIPS2019_9015}. As shown in Fig.~\ref{fig:loss}, the training and test losses are closely matched after $\sim 200$ epochs, indicating no overfitting.

After training, {\sc{Dingo}} can rapidly generate approximate posterior samples for any observed data. However, the results may deviate from the true posterior due to 
epistemic uncertainty, which arises when the neural network is either undertrained or not flexible enough \citep{Kendall:2017tnb}.

A practical way to check and refine results is through importance sampling \citep[IS,][]{https://doi.org/10.1002/wics.56, Elvira_2021}.  Combining this method with {\sc Dingo} leads to {\sc Dingo-IS} \citep{2023PhRvL.130q1403D}. It works by assigning weights to a set of $n$ samples $\theta_i$ drawn from the distribution $q(\theta|d)$. The weight for each sample is calculated as follows:  
\begin{equation}
\label{eq:weights}
w_i = \frac{\mathcal{L}(d|\theta_i)\pi(\theta_i)}{q(\theta_i|d)},
\end{equation}
where $\mathcal{L}(d|\theta_i)$ is the likelihood function, as defined in Eq.~\ref{eq:likelihood}. In an ideal case, all the weights $w_i$ would be equal. However, in practice, the number of effective samples depends on the variance of the weights \citep{Kong}, and it is given by:

\begin{equation}
\label{eq:neff}
n_{\mathrm{eff}} = \frac{\left( \sum_i w_i \right)^2}{\sum_i w_i^2}\,.
\end{equation}

The sample efficiency serves as a measure of how well the distribution $q(\theta|d)$ matches the target, and it is defined as:
\begin{equation}
\epsilon= \frac{n_{\mathrm{eff}}}{n} \in (0,100 \%].
\end{equation}
The Bayesian evidence $\mathcal{Z}(d)$ in Eq.~\ref{eq:bayes} and its uncertainty can be estimated by averaging the weights, as follows \citep{mcbook}:
\begin{equation}
    \mathcal{Z}(d) \pm \sigma = \frac{1}{n} \sum_i w_i \left ( 1 \pm \sqrt{\frac{1-\epsilon}{n-\epsilon}}\right)\,.
\end{equation}
The evidence is not only useful for model comparison, as in the standard Bayesian analysis, but can also be used to further validate NPE results. For example, the evidence would be biased if the support of the target posterior were not fully covered by $q(\theta|d)$ \citep{mcbook,2023PhRvL.130q1403D}.


\subsection{{\sc{GWFish+Priors}}}
\label{sec:fisher}

The FIM formalism is commonly used to simulate the PE performance of 3G detectors \citep{Cutler:1994ys,2008PhRvD..77d2001V,Chan:2018csa,Grimm:2020ivq,Borhanian:2020ypi,Iacovelli:2022bbs,Branchesi:2023mws}. In this framework, the likelihood in Eq.~\ref{eq:bayes} is approximated as a multivariate Gaussian distribution centered at given injected parameters $\theta^{\mathrm{inj}}$:
\begin{equation}
    \label{eq:fimlikelihood}
    \mathcal{L}(d|\theta) \propto \exp \left( -\frac{1}{2} \left(\theta - \theta^{\mathrm{inj}}\right)^\mathrm{T}~\mathcal{F}~ \left(\theta - \theta^{\mathrm{inj}}\right) \right),
\end{equation}
where $\mathcal{F}$ is the FIM:
\begin{equation}
\label{eq:fisher}
    \mathcal{F}_{ij} = \left( \frac{\partial h(\theta)}{\partial \theta_i} \Bigg | \frac{\partial h(\theta)}{\partial \theta_j} \right) \Bigg |_{\theta=\theta^{\mathrm{inj}}},
\end{equation}
with the noise-weighted inner product $(...|...)$ defined as in Eq.~\ref{eq:inner}.


We use the implementation of the FIM formalism provided in 
{\sc GWFish+Priors} \citep{Dupletsa:2024gfl}, where posterior samples randomly drawn from Eq.~\ref{eq:fimlikelihood} are weighted proportionally to the chosen prior distribution in Table~\ref{tab:prior}. 

The noise-weighted inner product $(...|...)$, defined in Eq.~\ref{eq:inner}, is used to compute the optimal SNR:  
\begin{equation}  
\label{eq:optimalSNR}
    \rho = \sqrt{(h(\theta^{\mathrm{inj}})|h(\theta^{\mathrm{inj}}))}.  
\end{equation}  
In a detector network, as in our case, the combined optimal SNR is obtained by summing the individual detector SNRs in quadrature.  


\section{Results}

\subsection{Single event}
\label{sec:single}

\begin{figure}
    \centering
    \includegraphics[width=0.9\linewidth]{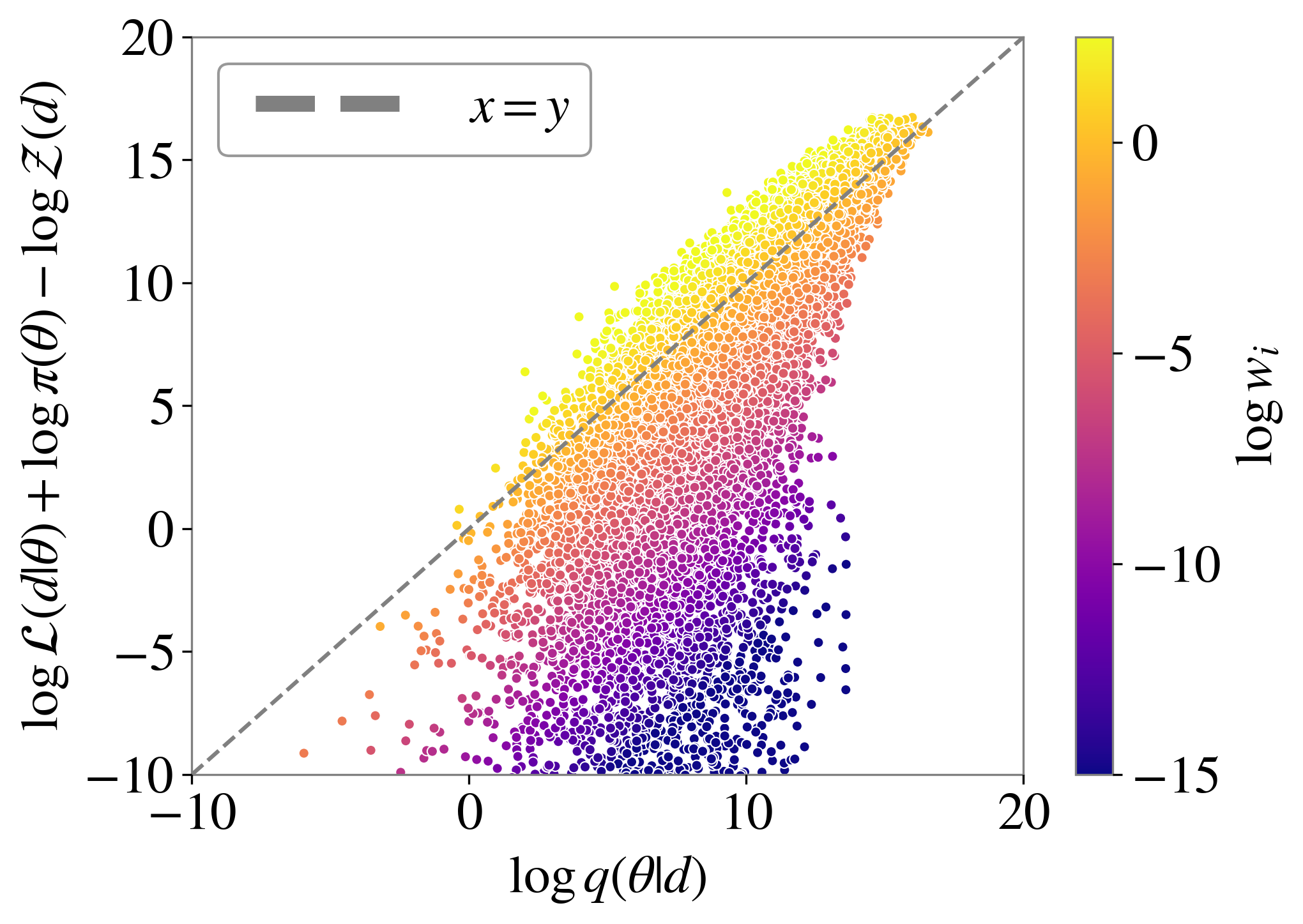}
    \caption{For each posterior sample drawn for \texttt{Event 1}, we compare the inferred density $\log q(\theta | d)$ with the normalized posterior $\log \mathcal{L}(d | \theta) + \log \pi(\theta) - \log \mathcal{Z}(d)$. The difference between these log densities determines the importance weights (Eq.~\ref{eq:weights}), which are color-coded. The sample efficiency is $\sim 16\%$. See Sec.~\ref{sec:dingo} for details.}
    \label{fig:weights}
\end{figure}

\begin{table}[]


    \centering
    \begin{tabular}{llccl}
    \toprule
        Parameter & $\theta^{\mathrm{inj}}$ & $\langle\mathrm{JSD}\rangle$ & JSD$^{\mathrm{thr}}$ & $\theta$ \\
        \hline
         &  & & \\   
        $\mathcal{M}_{\mathrm{d}}$ & 503.7 M$_\odot$ &  
        0.6 & 1.7 & ${516}_{-19}^{+16}$ M$_\odot$ \\
        $q$ & 0.43 &  
        0.9 & 2.1 & ${0.47}_{-0.04}^{+0.04}$ \\
        $d_{\mathrm{L}}$ & 20499.4 Mpc &  
        0.9 & 2.0 & ${20000}_{-3000}^{+2000}$ Mpc  \\
        ra & 2.12 rad &  
        0.7 & 1.7 & ${2.43}_{-1.82}^{+3.10}$ rad \\
        dec &  -1.17 rad & 
        0.7 & 1.5 & ${-0.03}_{-1.16}^{+1.22}$ rad \\
        $\theta_{\mathrm{jn}}$ & 2.21 rad &  
        0.7 & 1.4  & ${1.28}_{-0.33}^{+0.91}$ rad\\
        $\phi_c$ & 3.36 rad &  
        0.4 & 1.4 & ${3.76}_{-0.33}^{+0.26}$ rad \\
        $\psi$ & 2.18 rad  & 
        1.1 & 2.5 & ${1.59}_{-0.98}^{+0.95}$ rad\\
        $t_{\mathrm{geocent}}$ & 0.00005 s  & 
        0.1 & 0.5 & ${0.015}_{-0.011}^{+0.012}$ s\\
        $\chi_1$ & 0.38 &  
        0.3 & 1.2 & ${0.39}_{-0.09}^{+0.09}$ \\   
        $\chi_2$ & -0.39  & 
        0.9 & 2.7 & ${-0.12}_{-0.29}^{+0.18}$ \\   
         &  & & \\   
        $m_{1,\mathrm{s}}$ & 258.9 M$_\odot$ & & & $255_{-12}^{+18}$ M$_\odot$\\
        $m_{2,\mathrm{s}}$ & 111.8 M$_\odot$ & & & $121_{-13}^{+16}$ M$_\odot$\\
        $z$ & 2.5 & & & $2.4_{-0.3}^{+0.2}$\\
        $\rho$ & 83 &\\
        $\Delta \Omega_{90\%}$ & & & & 449 $\mathrm{deg}^2$ \\
    \bottomrule
    \end{tabular}
    \caption{Injected parameters $\theta^{\mathrm{inj}}$ for \texttt{Event 1}, along with the median Jensen-Shannon divergence ($\langle\mathrm{JSD}\rangle$, in units of $10^{-3}$ nat), which quantifies the deviation between {\sc Dingo-IS} and {\sc Bilby} for one-dimensional marginal posteriors. For comparison, the corresponding JSD threshold (JSD$^{\mathrm{thr}}$) is also shown. The last column on the right shows the median and 90\% credible interval of the recovered parameters $\theta$ with {\sc Dingo-IS}. We report in the bottom rows the source-frame primary and secondary mass, redshift, the optimal signal-to-noise ratio ($\rho$, see Eq.~\ref{eq:optimalSNR}), and the sky localization error. See Sec.~\ref{sec:single} for details.}
    \label{tab:jsd}
\end{table}

\begin{figure*}
    \centering
    \includegraphics[width=1\linewidth]{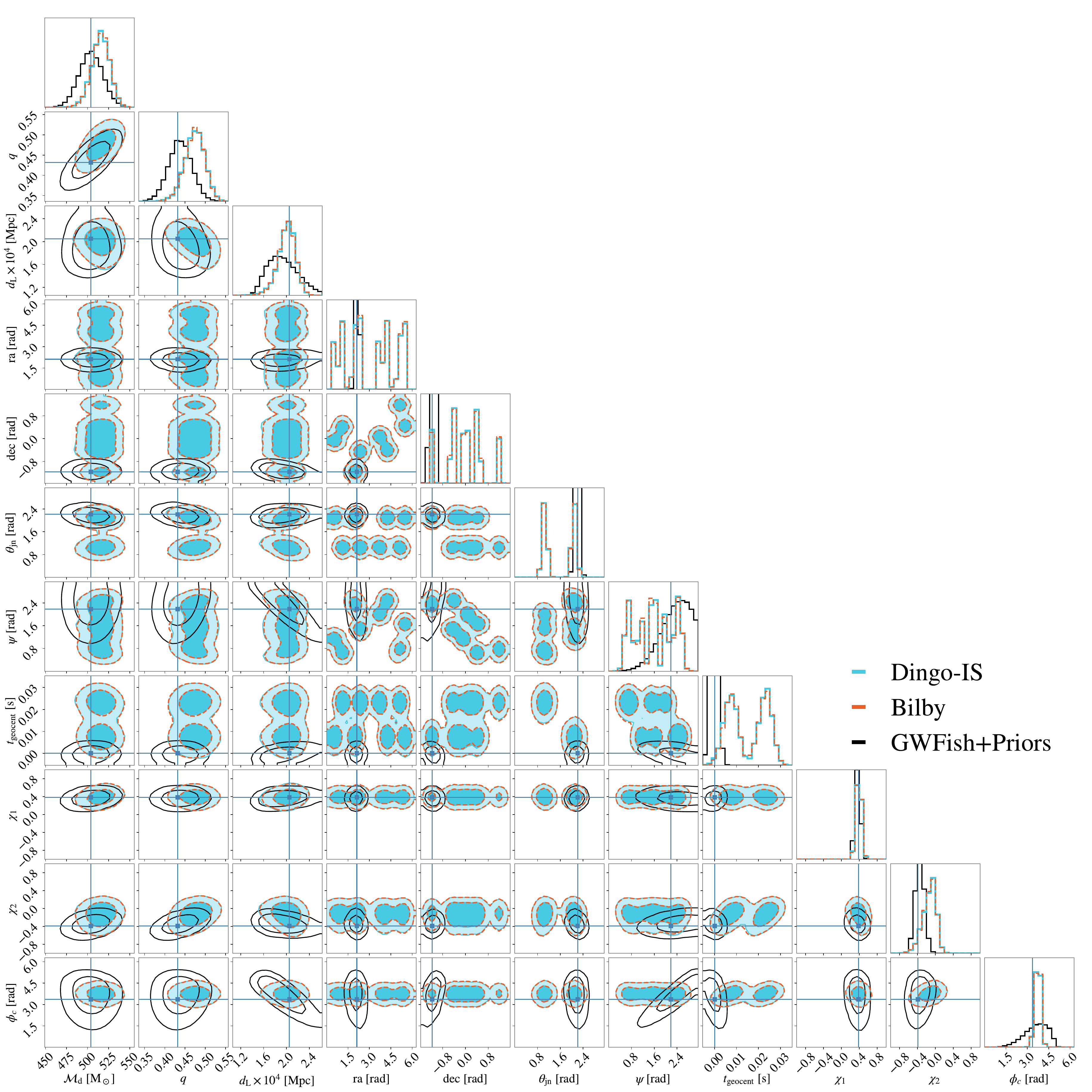}
    \caption{Marginalized one- and two-dimensional posterior distributions for \texttt{Event 1} over all sets of parameters, comparing {\sc Dingo-IS} (light blue), {\sc Bilby} (orange), and {\sc GWFish+Priors} (black). Vertical and horizontal lines mark the true injected values (see Table~\ref{tab:jsd}). Contours represent 68\% and 95\% credible regions. Sky modes are discussed in Sec.~\ref{sec:skyloc}. See Sec.~\ref{sec:single} for other details.}
    \label{fig:bilby}
\end{figure*}

We first validate {\sc Dingo-IS}, trained on ET-$\Delta$, by comparing its results with those obtained using the standard inference code {\sc Bilby} for a randomly selected event from the prior distribution, referred to hereafter as \texttt{Event 1}. The injected and recovered parameters for this event are listed in Table~\ref{tab:jsd}, and the corresponding signal and noise realizations are shown in Fig.~\ref{fig:data}. A second event, \texttt{Event 2}, with a higher injected redshift and lower optimal SNR, is analyzed in Appendix~\ref{app:event2}, leading to similar conclusions.

For this injection, we draw $3 \times 10^4$ samples from the proposal distribution $q(\theta|d)$. We then analytically sample the phase $\phi_c$ and evaluate the importance weights (Eq.~\ref{eq:weights}), which are shown in Fig.~\ref{fig:weights}. The sample efficiency of this event is equal to $\sim 16\%$, corresponding to $5 \times 10^3$ effective samples. 

From the importance weights shown in Fig.~\ref{fig:weights}, we observe a tail with high $q(\theta|d)$ and low $p(\theta|d)$, without a corresponding tail with high $p(\theta|d)$ and low $q(\theta|d)$. This suggests that the {\sc Dingo} proposal $q(\theta|d)$ covers the full support of the posterior distribution and slightly overestimates its width, which is a desired characteristic of NPE. This is because an overly broad posterior approximation results in a much smaller loss of sample efficiency compared to one that is too narrow \citep{2023PhRvL.130q1403D}.

In {\sc Bilby} we set the number of live points to \texttt{nlive = 5000}. The log evidence calculated using {\sc Dingo-IS} and {\sc Bilby} is $\log \mathcal{Z}(d) = -6073.89 \pm 0.01$ and $-6073.58 \pm 0.08$, respectively. This small difference is within statistical uncertainty.

We compare posterior samples from {\sc Dingo-IS} and {\sc Bilby} using the Jensen–Shannon divergence \citep[JSD;][]{61115}, which is a symmetric extension of the Kullback–Leibler divergence \citep{10.1214/aoms/1177729694}. JSD quantifies the difference between two probability distributions, on a scale from 0 nat (identical distributions) to ln(2) = 0.69 nat (maximally distinct distributions), where the nat unit arises from using the natural logarithm. 

To establish a threshold for the JSD above which one-dimensional posteriors are considered to be statistically different, we proceed as follows. We use {\sc{Dingo-IS}} to generate 100 sets of $10^4$ posterior samples. For each parameter $\theta$, we calculate the JSD for all 4950 unique pairs of these sets. Since all these samples are drawn from the same distribution, the resulting JSD values reflect statistical fluctuations rather than meaningful differences. We then take the 95th percentile of the resulting JSD values as a threshold for each parameter (JSD$^{\mathrm{thr}}$ in Table~\ref{tab:jsd}). Finally, we compare the one-dimensional posteriors obtained with {\sc{Bilby}} to each of the {\sc{Dingo-IS}} sets, and we check whether the median of these JSD values is below the corresponding threshold ($\langle \mathrm{JSD}\rangle$ in Table~\ref{tab:jsd}).

As shown in Table~\ref{tab:jsd}, the evaluated JSDs over marginal one-dimensional posteriors indicate that the two distributions are effectively identical. This perfect agreement is further illustrated in Fig.~\ref{fig:bilby}, which presents the marginalized one- and two-dimensional posteriors for all parameters.

Figure~\ref{fig:bilby} shows the multimodalities and degeneracies that arise due to the geometry of a triangular colocated detector \citep{Singh:2020wsy,Singh:2021bwn}. These effects appear in parameters such as the right ascension (ra), declination (dec), inclination angle ($\theta_{\mathrm{jn}}$), polarization angle ($\psi$) and geocentric time ($t_{\mathrm{geocent}}$, see also Sec.~\ref{sec:geocentric-time}). In particular, Sec.~\ref{sec:skyloc} discusses the degeneracies in sky localization, which lead to eight distinct sky modes in the two-dimensional marginal distribution of ra and dec. These modes are absent by construction when using the FIM approximation for PE. 

Moreover, the posteriors obtained with {\sc Dingo-IS} and {\sc Bilby} are slightly offset from the injected value due to the influence of noise in the data, whereas the FIM uncertainties, which do not account for the specific data realization, remain centered on the true value.

\subsection{Population analysis}
\label{sec:pop}

\begin{figure}
    \centering
    \includegraphics[width=0.9\linewidth]{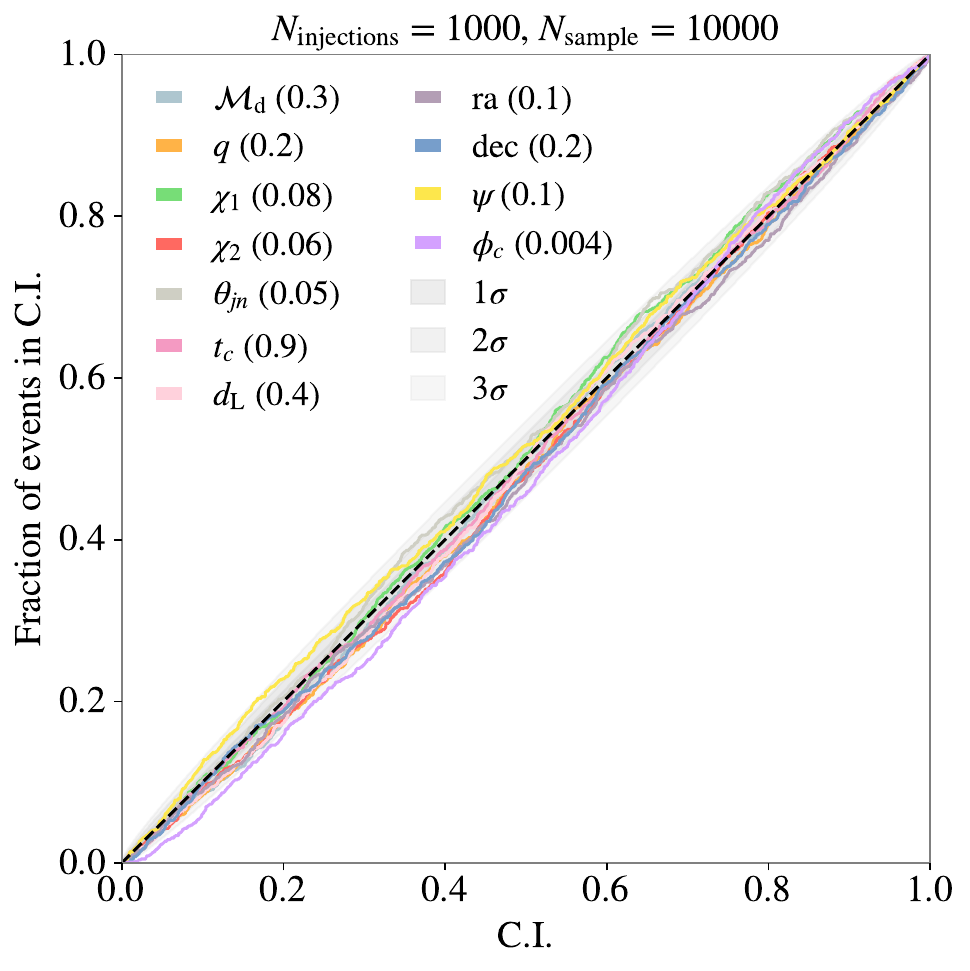}
    \caption{Probability--probability (P--P) plot showing the confidence interval (C.I.) on the $x$-axis against the fraction of events within that C.I on the $y$-axis, for posterior distributions obtained with {\sc Dingo} (without importance sampling) from 1000 injected binary black holes (BBHs) signals sampled from the prior (Table~\ref{tab:prior}). The shaded regions represent the $1\sigma$, $2\sigma$, and $3\sigma$ C.I., with $p$-values given for each parameter in parenthesis and a combined $p$-value
of 0.001. See Sec.~\ref{sec:pop} for details.}
    \label{fig:pp_plot}
\end{figure}

\begin{figure}
    \centering
    \includegraphics[width=1\linewidth]{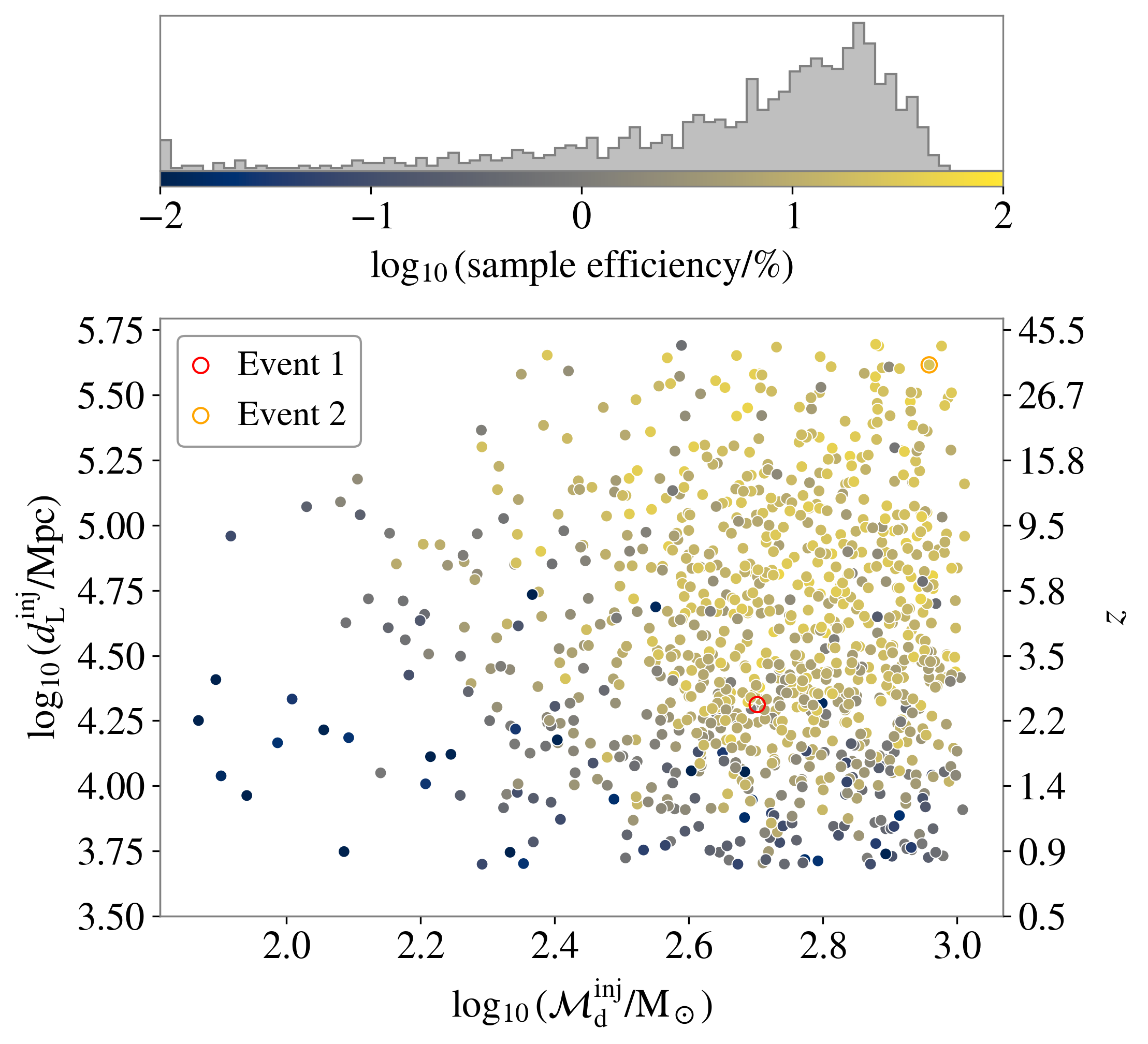}
    \caption{Injections processed with {\sc Dingo-IS}, shown as a function of detector-frame chirp mass ($x$-axis) and luminosity distance (left $y$-axis), with corresponding redshift values (right $y$-axis). Each point is color-coded by the sample efficiency, and its marginal distribution is shown as a gray histogram above. \texttt{Event 1} and \texttt{Event 2}, discussed in Sec.~\ref{sec:single} and Appendix~\ref{app:event2}, are highlighted with red and orange circles, respectively. See Sec.~\ref{sec:pop} for further details.}
    \label{fig:sample_efficiency}
\end{figure}

We generate 1000 simulated BBH signals with parameters randomly sampled from the prior distribution (Table~\ref{tab:prior}) and process them using {\sc Dingo-IS}. For each injection, we draw $10^4$ samples, estimate the phase $\phi_c$ in post-processing, and compute the importance weights, as detailed in Sec.~\ref{sec:dingo}. 

Fig.~\ref{fig:pp_plot} presents a probability--probability (P--P) plot, which is a standard method to validate inference algorithms \citep{Cook01092006,Veitch:2014wba,talts2020validatingbayesianinferencealgorithms,2020PhRvD.102j4057G}. The P-P plot checks whether the fraction of injected values recovered within a given confidence interval follows a uniform distribution. This also allows us to compute the $p$-values for each parameter. For an ideal sampler, an $n\%$ confidence interval should contain $n\%$ of the events on average, making a diagonal line in the plot. Statistical fluctuations cause deviations, so we also display the $1\sigma$, $2\sigma$, and $3\sigma$ confidence intervals. 

Figure~\ref{fig:pp_plot} confirms that the {\sc Dingo} proposal distribution $q(\theta|d)$, trained for ET-$\Delta$, performs reasonably well for most parameters, with the curves following the diagonal. However, some deviations are visible, most notably for the phase $\phi_c$, as also indicated by its respective $p$-value. This makes importance sampling necessary to correct deviations in samples drawn with NPE.

Fig.~\ref{fig:sample_efficiency} shows how the sample efficiency varies with detector-frame chirp mass and luminosity distance of the processed injections. {\sc Dingo-IS} enables PE in a wide redshift range ($1 \lesssim z \lesssim 45$) with high accuracy. The average and median sample efficiency are 12.5\% and 10\%, respectively, with $\sim 85$\% of sources with sample efficiency $> 1\%$. The highest sample efficiencies (about 10\%--20\%) are observed for the most distant BBHs. They have low SNR, making them easier for NPE to learn, as their posteriors are closer to the prior. This can be also seen as the average sample efficiency increases to $\sim 18\%$ $(\sim 20 \%)$ if we consider only sources merging at $z > 4$
$(z > 10)$. Similar sample efficiencies were reported in \cite{2023PhRvL.130q1403D}, where they analyzed 42 BBH events detected with the LIGO and Virgo interferometers \citep{LIGOScientific:2020ibl,KAGRA:2021vkt}.

Fig.~\ref{fig:sample_efficiency} also highlights regions of low sample efficiency, particularly at $z \lesssim 2$. This behavior might be influenced by the chosen prior on the luminosity distance. As shown in previous studies \citep[e.g.,][]{Green:2020dnx}, using a uniform prior in luminosity distance might improve performance by exposing the neural network to more low-$d_{\mathrm{L}}$ events during training. In regions with low sample efficiency, we can compensate by increasing the number of posterior samples drawn with {\sc Dingo-IS}.

Our goal is to ensure that each processed injection has enough effective samples, around $10^3$. Therefore, for the remainder of the manuscript, we discard sources with a sample efficiency below 10\%, leaving a total of 
500 sources. As shown in Fig.~\ref{fig:sample_efficiency}, this selection primarily retains high-redshift events, in line with the objectives of this study.

\begin{figure}
    \centering
    \includegraphics[width=0.9\linewidth]{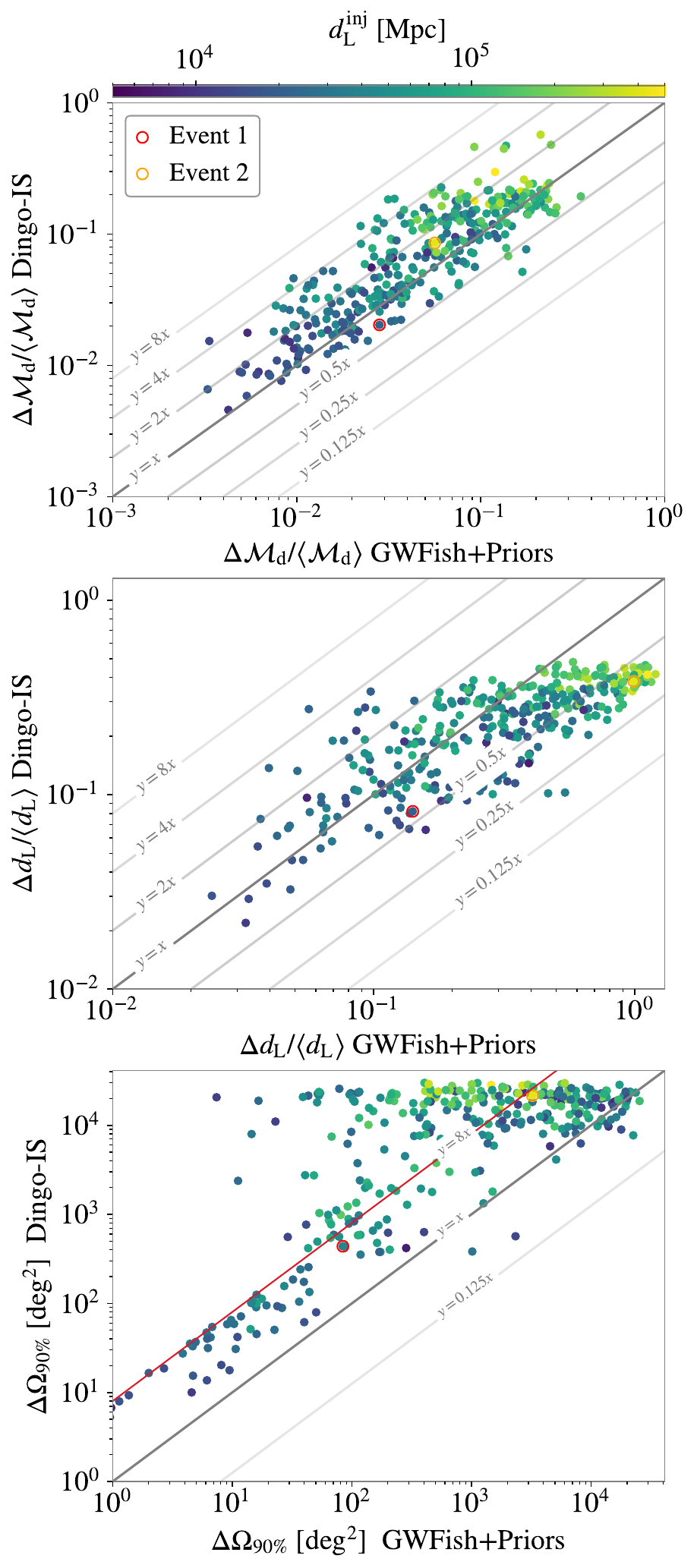}
    \caption{The top, middle, and bottom panels show the relative variations for the detector-frame chirp mass ($\mathcal{M}_{\mathrm{d}}$), luminosity distance ($d_{\mathrm{L}}$), and sky localization ($\Delta \Omega_{90\%}$), respectively. The $x$-axis corresponds to results from Fisher-information matrix approximation (FIM, {\sc GWFish+Priors}), while the $y$-axis represents neural posterior estimate with importance sampling (NPE, {\sc Dingo-IS}). The points are color-coded by the injected luminosity distance. \texttt{Event 1} (\texttt{Event 2}) discussed in Sec.~\ref{sec:single} (Appendix~\ref{app:event2}) is marked with a red (orange) circle. See Sec.~\ref{sec:pop} for details.}
    \label{fig:scatter}
\end{figure}

\begin{figure}
    \centering
    \includegraphics[width=0.9\linewidth]{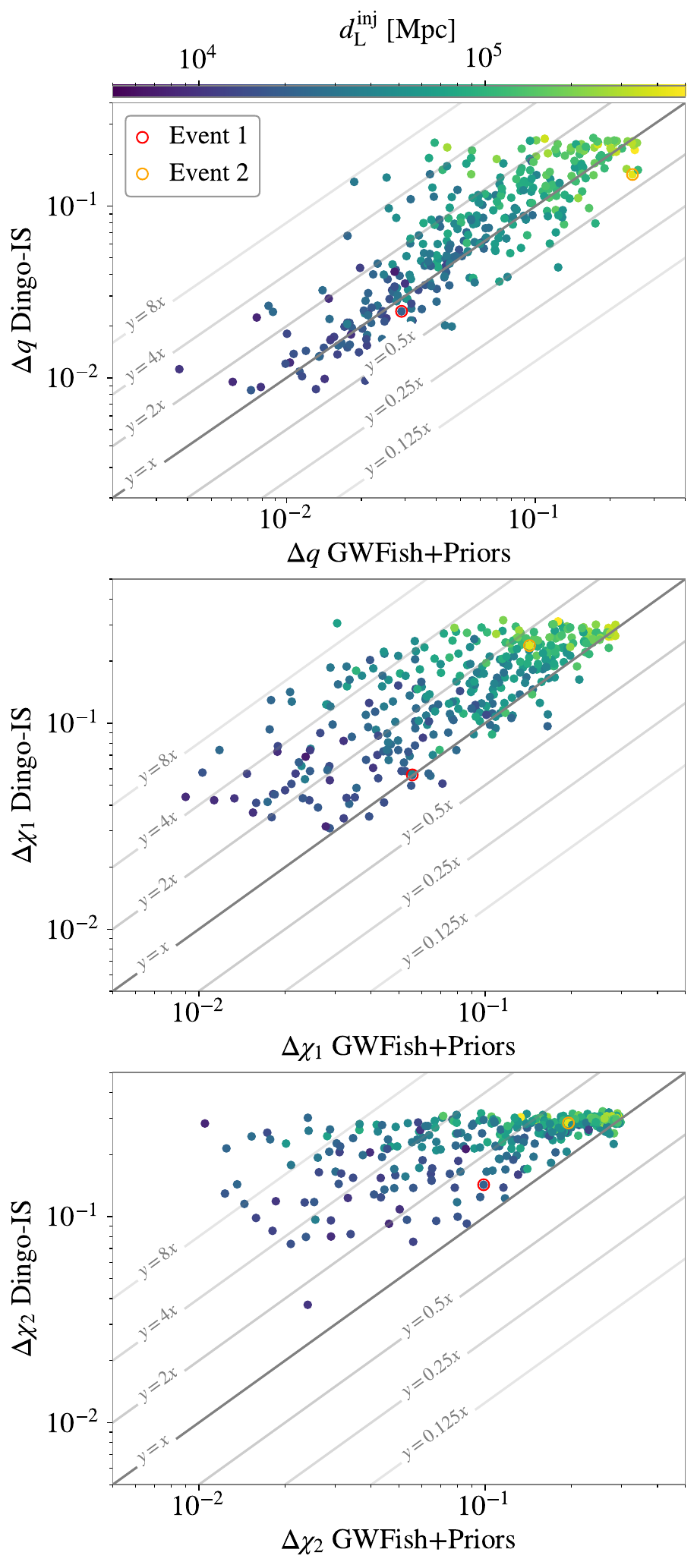}
    \caption{Same as Fig.~\ref{fig:scatter}, but with the top, middle, and bottom panels showing the errors for the mass ratio ($q$) and the first and second aligned spins ($\chi_1$, $\chi_2$), respectively.}
    \label{fig:scatter_spins}
\end{figure}

Figure~\ref{fig:scatter} presents scatter plots that compare the relative variations in chirp mass, luminosity distance, and sky localization estimated with NPE and FIM. In particular, we compute the 90\% highest probability density sky area by evaluating a kernel density estimate on a multiorder HEALPix grid \citep{Gorski:1999rt}, using the \texttt{ligo-skymap} library \citep{Singer:2015ema,Singer:2016eax,Singer:2016erz}. The other reported intervals, such as $\Delta x/\langle x\rangle$, represent the relative variation of the parameter around its average, where $\Delta x$ is the square root of the corresponding covariance element, reflecting the $1\sigma$ uncertainty.

For the chirp mass, FIM systematically underestimates the error compared to NPE. In the worst cases, the difference remains within a factor of 4 and is independent of the injected luminosity distance. 

For sources with injected luminosity distance up to $d^{\mathrm{inj}}_\mathrm{L} \sim 10^5~\mathrm{Mpc}$, the FIM overestimates and underestimates the relative variations compared to NPE by up to a factor of 4. Moreover, for $d^{\mathrm{inj}}_\mathrm{L} > 10^5$ Mpc, the estimated relative variation from \textsc{GWFish+Priors} increases systematically, reaching $\sim 1$. This indicates that the FIM analysis returns the chosen prior range (Table~\ref{tab:prior}). The relative variation on luminosity distance is on average between 2 and 4 times larger compared to what is derived with NPE. Therefore, FIM tends to overestimate luminosity distance errors for high-redshift sources. The accurate evaluation of distance errors is important for analyses of the ET science case, as discussed in Sec.~\ref{sec:impli}.

For sky localization, the discrepancy is the most severe. As seen in Fig.~\ref{fig:bilby}, ET-$\Delta$ sky localization for short-lived sources is characterized by eight degenerate sky modes (see Sec.~\ref{sec:skyloc} for a discussion), which FIM fails to capture by construction. For well-localized sources ($\Delta \Omega_{90\%}<100~\mathrm{deg}^2$), the discrepancy appears to be very close to a factor of 8. For poorly localized sources ($\Delta \Omega_{90\%}>100~\mathrm{deg}^2$), the error difference can increase by up to 3 orders of magnitude, worsening with the injected distance. This has also important implications for the ET science case, discussed in Sec.~\ref{sec:impli}. 

Figure~\ref{fig:scatter_spins} shows that the FIM analysis tends to underestimate the error for the mass ratio (within a factor of 4) and systematically underestimates the errors for the aligned spins. In particular, the uncertainty on $\chi_1$ and $\chi_2$ for most sources is $\Delta \chi_i \sim 0.3$, which is the spread of the prior distribution (Eq.~\ref{eq:spins}). This indicates that aligned spins estimated with NPE are mostly unconstrained for these sources. These findings are consistent with previous studies, such as \cite{Dupletsa:2024gfl}.

Figures~\ref{fig:scatter} and \ref{fig:scatter_spins} display 344 of the 500 selected events that have a sample efficiency above 10\%. The remaining 156 events are excluded because their covariance matrices, given by the inverse of the FIM in Eq.~\ref{eq:fisher}, are close to singular. This exclusion is required in {\sc{GWFish+Priors}} to ensure numerical stability when sampling from a truncated multivariate Gaussian, as described in \cite{Dupletsa:2024gfl}.

Moreover, the results shown in Figs.~\ref{fig:scatter} and \ref{fig:scatter_spins} are based on quasi-circular systems with aligned spins, which means we do not account for spin-induced precession of the orbital plane or orbital eccentricity. If measured, precession and/or eccentricity could help break degeneracies between waveform parameters, leading to even more well-constrained posteriors \citep{Lower:2018seu,Yang:2022fgp,Yang:2022tig,Ng:2022vbz,Saini:2023wdk}. We discuss this further in Sec.~\ref{sec:hom}.


\section{Discussion}

\subsection{Sky modes}
\label{sec:skyloc}

\begin{figure}
    \centering
    \includegraphics[width=1\linewidth]{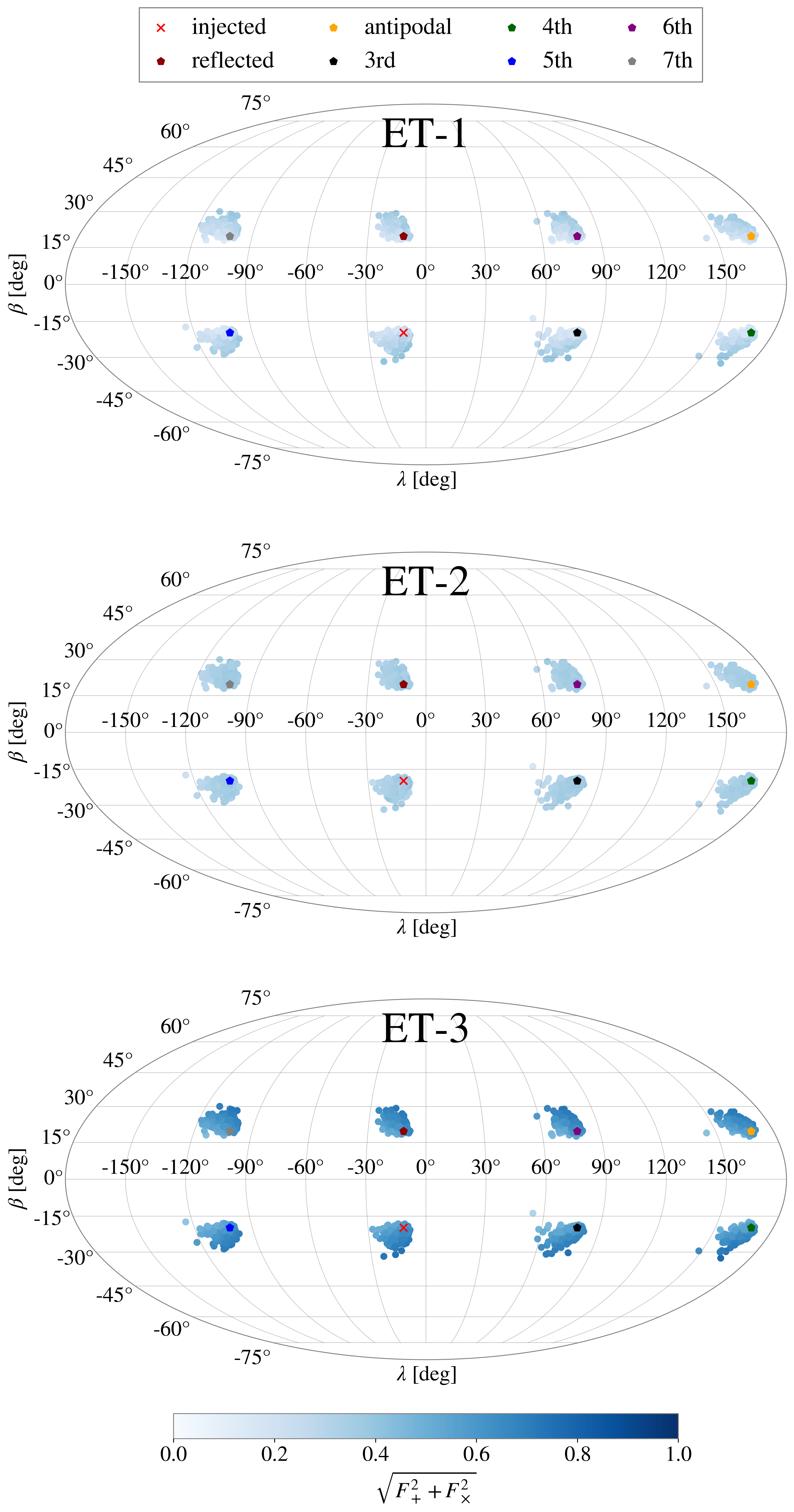} 
    \caption{Right ascension and declination samples from {\sc Dingo-IS} converted in azimuth ($\lambda$) and altitude ($\beta$). Degeneracies, as listed in Table~\ref{tab:sky_modes}, are included. Color-coded is the square root of the sum of the antenna power pattern functions (Eq.~\ref{eq:pattern}) of ET-1 (top panel), ET-2 (middle panel) and ET-3 (bottom panel). See Sec.~\ref{sec:skyloc} for details.}
    \label{fig:horizon}
\end{figure}

\begin{table}
    \centering
    \begin{tabular}{lc}
\toprule
Sky mode & Coordinates  \\
\hline
Injected & $\beta, \lambda$\\
Reflected & $-\beta, \lambda$  \\
Antipodal & $-\beta, \lambda + \pi$ \\
3rd            & $\beta, \lambda + \pi/2$ \\
4th           & $\beta, \lambda + \pi$  \\  
5th           & $\beta, \lambda - \pi/2$  \\  
6th           & $-\beta, \lambda + \pi/2$ \\  
7th           & $-\beta, \lambda - \pi/2$ \\ 
\bottomrule
    \end{tabular}
    \caption{Sky mode degeneracies reported in Fig.~\ref{fig:horizon}. We use the same nomenclature as in \cite{Marsat:2020rtl}. Coordinates are given in the horizontal reference frame, defined by altitude ($\beta$) and azimuth ($\lambda$) angles. See Sec.~\ref{sec:skyloc} for details. }
    \label{tab:sky_modes}
\end{table}

    
The degeneracies in sky localization, which are evident in the two-dimensional marginal distribution of ra and dec in Fig.~\ref{fig:bilby}, reveal eight distinct sky modes. These modes have been previously discussed for ET in its triangular configuration (see Fig.~2 in \citealt{Singh:2021bwn} and Fig.~2 in \citealt{Singh:2020wsy}) and in the LISA triangular space-based interferometer, assuming frequency- and time-independent antenna patterns \citep{Baibhav:2020tma,Marsat:2020rtl}. For consistency, we label these modes in Table~\ref{tab:sky_modes} using the same nomenclature as in \cite{Marsat:2020rtl}.

These sky modes result from the symmetries in the antenna pattern functions of the three detectors that form ET-$\Delta$. The antenna response functions, which take into account that GW detectors are not equally sensitive to gravitational waves from every direction, can be expressed in the horizontal reference frame, defined by the azimuth ($\beta$) and altitude ($\lambda$) angles. In this frame, the antenna amplitude pattern functions are given by \citep{Regimbau:2012ir}:

\begin{multline}
\label{eq:fplus}
        F^{i}_{+}(\beta, \lambda, \psi) = -\frac{\sqrt{3}}{4}[(1+\cos^2\beta)\sin2\lambda\cos2\psi + \\ 2\cos\beta\cos 2\lambda\sin 2\psi ],
\end{multline}

\begin{multline}
\label{eq:fcross}
        F^{i}_{\times}(\beta, \lambda, \psi) = +\frac{\sqrt{3}}{4}[(1+\cos^2\beta)\sin2\lambda\sin2\psi - \\ 2\cos\beta\cos 2\lambda\cos 2\psi ],
\end{multline}
where $i =$ ET-1. The antenna amplitude pattern functions of ET-2 and ET-3 are obtained from $F^{i}_{+}$ and $F^{i}_{\times}$ by the transformation $\lambda \to \lambda \pm 2\pi/3$. 


All these antenna patterns are symmetric under the following transformations: reflection of the sky position across the horizontal plane (\( \beta \to -\beta \)), rotation about the vertical direction by half a turn (\( \lambda \to \lambda + \pi \)), and simultaneous rotation of the sky position and polarization angle by a quarter of a turn (\( \lambda \to \lambda + \pi/2 \), \( \psi \to \psi + \pi/2 \)) \citep{Marsat:2020rtl}. Because these transformations produce identical signals in each ET component, they lead to eight degenerate sky positions in the posterior, evenly spaced in \( \lambda \) and symmetrically distributed above and below the horizon. 

To visualize this effect, we show the total response of each ET component to an incident wave, also called antenna power pattern function: for ET-1, this is given by \citep{Schutz:2011tw}:
\begin{align}
\label{eq:pattern}
    \sqrt{(F^i_+)^2 + (F^i_\times)^2} = \notag \\ \sqrt{\frac{3}{12} \left[ (1+\cos^2\beta)^2 \sin^2 2\lambda + \cos^2\beta\cos^2 2\lambda \right]}    
\end{align}
where the dependency on the polarization angle $\psi$ cancels out. Figure~\ref{fig:horizon} displays the sky location samples in the horizontal reference frame, color-coded by Eq.~\ref{eq:pattern}. The antenna responses remain identical across all eight modes in all three ET detectors.

The same degeneracies also occur when analyzing the PE of massive black hole binaries (MBHBs) with LISA \citep{Marsat:2020rtl,Pitte:2023ltw,Sun:2025cbo,Yi:2025pxe}, provided that the antenna pattern does not change due to Earth's rotation and the long-wavelength approximation is assumed \citep{Essick:2017wyl}. However, in LISA, these approximations can be relaxed, allowing for resolution of these degeneracies \citep{Buscicchio:2021dph}. Specifically, an MBHB with a total mass greater than $10^6~\mathrm{M}_\odot$ can stay in the LISA frequency band for more than 40 hours \citep{Marsat:2020rtl}, meaning the rotation of the satellite constellation can have a significant effect.

The long-wavelength approximation (also known as the low-frequency limit) assumes that the antenna pattern functions (Eq.~\ref{eq:fplus} and Eq.~\ref{eq:fcross}) act purely as projections of the GW signal onto the detector. This approximation is valid when the length of the detector arms is much shorter than the wavelength of the incoming GW \citep{Misner:1973prb,1989thyg.book.....H,Rakhmanov:2008is}. In reality, this approximation can break down, and the response can also depend on the GW frequency. The frequency at which the long-wavelength approximation becomes important is determined by the unperturbed roundtrip travel time along one arm, also known as the free spectral range \citep[FSR;][]{Marsat:2020rtl,Essick:2017wyl}:
\begin{equation}
    f_{\mathrm{FSR}} = \frac{c}{2L}
\end{equation}
Here, $c$ is the speed of light, and for LISA, $L=2.5\times 10^9~\mathrm{m}$, which gives $f_{\mathrm{FSR}} = 0.06$ Hz. The merger frequency of MBHBs within the LISA frequency band is of the same order of magnitude as $f_{\mathrm{FSR}}$, breaking the long-wavelength approximation \citep{Marsat:2020rtl}.

In comparison, for ET-$\Delta$, the free spectral range is
$f_{\mathrm{FSR}} = 1.5 \times 10^4~\mathrm{Hz}$, which is at least 3 orders of magnitude higher than the typical merger frequency of the sources considered in this study (see Eq.~\ref{eq:freq}). Moreover, BBH mergers with high detector-frame chirp mass last only a few seconds in the ET-$\Delta$ frequency band. Therefore, we conclude that assuming a time- and frequency-independent antenna pattern is still a valid approximation for high-redshift sources detected with ET-$\Delta$.

\subsection{Geocentric time}
\label{sec:geocentric-time}

\begin{figure}
    \centering
    \includegraphics[width=1\linewidth]{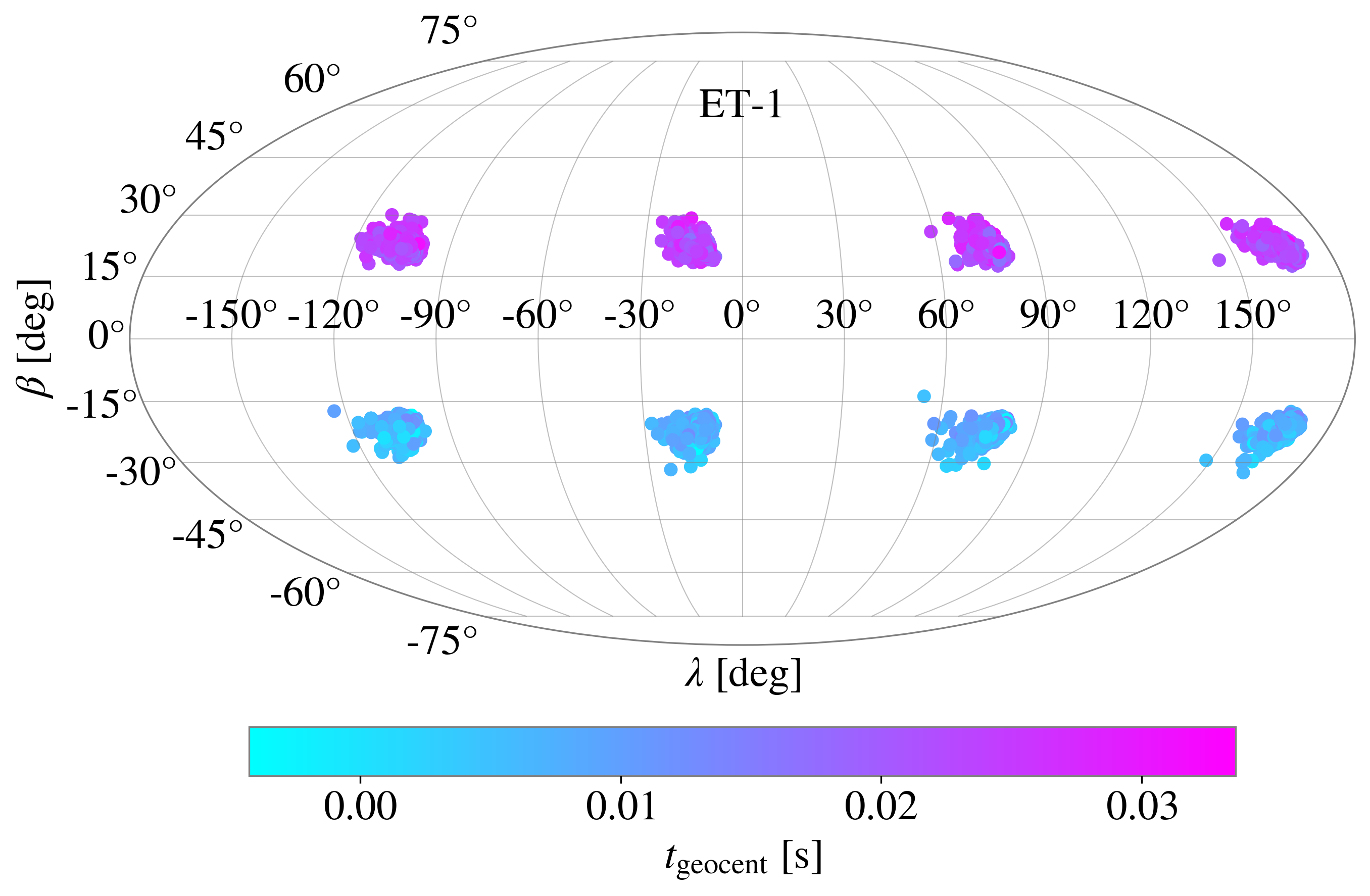} 
    \caption{Right ascension and declination samples from {\sc Dingo-IS} converted in azimuth ($\lambda$) and altitude ($\beta$). Color-coded is the geocentric time. See Sec.~\ref{sec:geocentric-time} for details.}
    \label{fig:geocent}
\end{figure}

The sky localization modes discussed in Sec.~\ref{sec:skyloc} also lead to a bimodal distribution in the geocentric time. This time is calculated by adding a delay to the arrival time at the detector site $(t_\text{det})$. This delay depends on the source position in the sky \citep{Singer:2015ema, Harry:2011gre}:
\begin{equation}
\label{eq:time-delay}
t _\text{geocent} = t _\text{det} + \frac{\hat{n}\cdot \vec{r}_\text{ET}}{c},
\end{equation}
where $\hat{n}$ is a unit vector pointing to the source in the sky and $\vec{r}_\text{ET}$ is the vector from the center of the Earth to the ET-$\Delta$ location.

The scalar product in Eq.~\ref{eq:time-delay} depends on the altitude angle $\beta$, defined as $\beta = \arccos(\hat{n} \cdot \hat{r}_\text{ET})$, where $\hat{r}_\text{ET}$ is the unit vector associated with $\vec{r}_\text{ET}$. Of the eight modes described in Sec.~\ref{sec:skyloc}, four are centered around the injected value of $\beta$, and the other four are centered around the opposite angle, $-\beta$, as shown in Figs.~\ref{fig:horizon} and \ref{fig:geocent}.

As a result, the geocentric time $t_\text{geocent}$ typically shows a bimodal distribution (see Fig.~\ref{fig:geocent}). The two modes are most separated when the source is near the local zenith or nadir ($|\beta| \approx \pi/2$). In contrast, when the source is near the local horizon ($\beta \approx 0$), the distribution becomes nearly unimodal (as \texttt{Event 2} in Appendix~\ref{app:event2}).

\subsection{Higher-order modes and precessing spins}
\label{sec:hom}

\begin{figure*}
    \centering
    \includegraphics[width=1\linewidth]{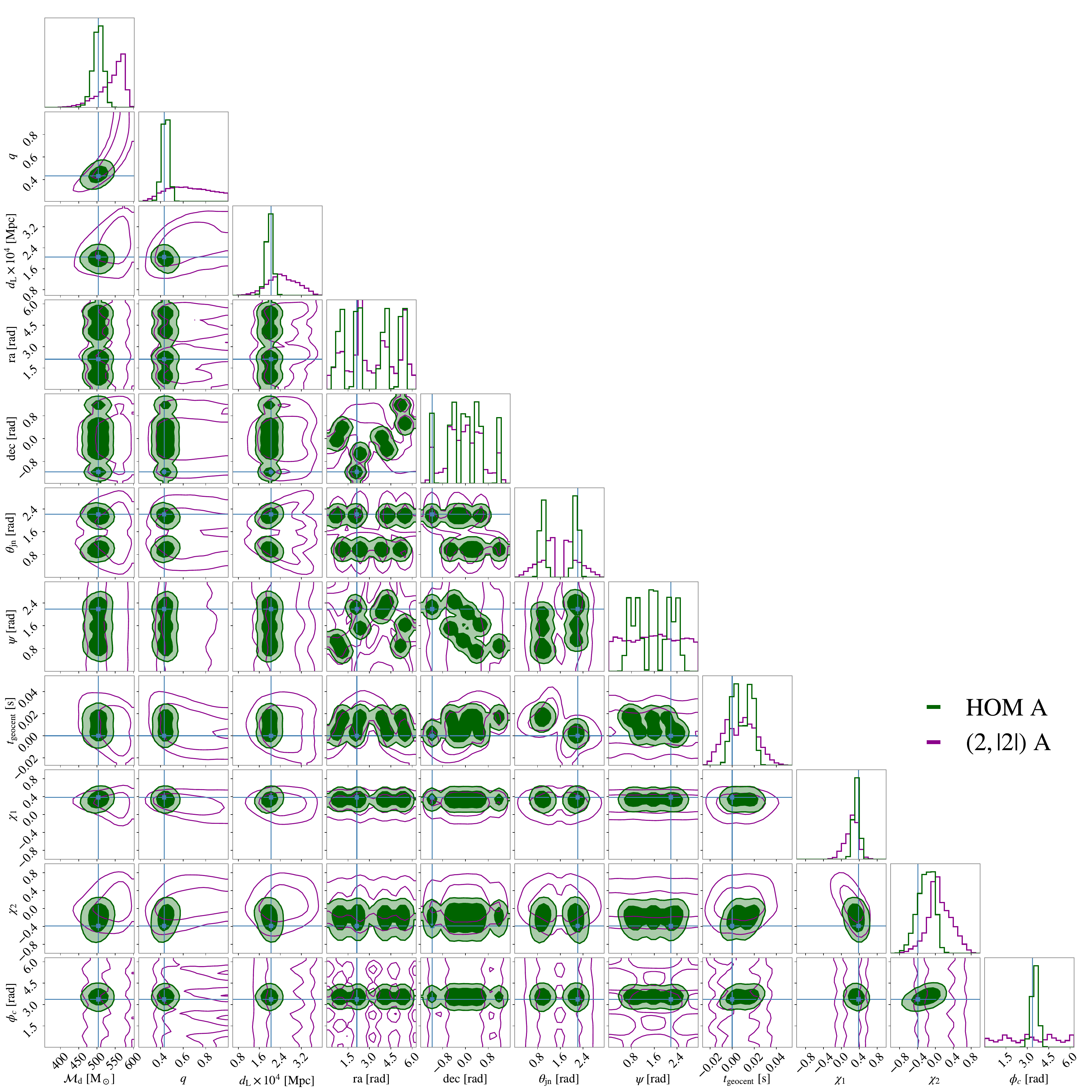}
    \caption{Marginalized one- and two-dimensional posterior distributions for \texttt{Event 1} with zero noise. We compare waveforms that include aligned spins and higher-order modes (HOM A, dark green) with those that include only the dominant mode ($(2,|2|)$ A, dark magenta), both analyzed using {\sc Bilby}. Vertical and horizontal lines indicate the true injected values (see Table~\ref{tab:jsd}), while the contours represent the 68\% and 95\% credible regions. Further details are provided in Sec.~\ref{sec:hom}.}
    \label{fig:hom}
\end{figure*}

In the time domain, a GW signal $h = h_+ + ih_\times$ can be expressed as \citep{Newman:1966ub,Berti:2007fi,Kidder:2007rt}:
\begin{equation}
    h(t; \vec{\lambda}) = \frac{1}{d_\mathrm{L}}\sum_{l=0}^\infty \sum_{m=-\ell}^{\ell} Y_{-2}^{\ell,m} ~ h_{\ell,m}(t;\vec{\lambda})
\end{equation}
where $Y_{-2}^{l,m}$ are the spin-weighted spherical harmonics of weight $-2$ \citep{Blanchet:2013haa,Mills:2020thr}, which capture the angular dependence of the radiation, and $\vec{\lambda}$ encompasses the intrinsic parameters, such as component masses and spins. The terms $h_{\ell,m}$ are the GW modes (or multipoles), with the dominant quadrupole mode corresponding to $\ell=2$, and $m=\pm2$. Higher-order modes (HOMs) appear for $\ell \geq 2$ and $|m| \neq 2$, and are also referred to as higher multipoles \citep{Khan:2019kot}, higher spherical harmonics \citep{Mills:2020thr}, nonquadrupole or subdominant modes.

In this work, we assume aligned spins, meaning the spins of the individual black holes are parallel to the orbital angular momentum of the binary system (Eq.~\ref{eq:aspins}), so no spin precession occurs \citep{Apostolatos:1994mx}. In this case, the waveform can be written in the frequency domain using the stationary-phase approximation, where the HOM amplitudes for both polarizations depend on the inclination angle \citep{Mills:2020thr}. 

When more than one mode is observed, the measurement of the inclination angle can be improved \citep{LIGOScientific:2020stg,Kalaghatgi:2019log}, reducing the correlation with the luminosity distance. In addition, detecting HOMs leads to a more accurate measurement of the properties of the source, in general, as they help to model the GW signal more accurately \citep{Kalaghatgi:2019log, LIGOScientific:2020stg,LIGOScientific:2020zkf,Marsat:2020rtl,Baral:2025geo,Yi:2025pxe,Pitte:2023ltw}. HOMs can aid in early warning detections since they have longer in-band signal duration in the time domain with respect to the dominant mode \citep{Kapadia:2020kss}. Lastly, they can be used to test general relativity \citep{Puecher:2022sfm}.

The contribution of HOMs relative to the leading mode depends largely on the system asymmetries. Their amplitude is higher in binaries with large mass ratios, making them particularly relevant for neutron star–black hole binaries \citep{Vitale:2018wlg,Zhang:2023ceh}, in binaries with large misaligned spins \citep{Singh:2023aqh}, and in sources viewed at large inclination angles, close to edge-on \citep[$\theta_{\mathrm{jn}} \sim \pi/2$;][]{Mills:2020thr}.

If spin precession is allowed, the symmetry \( h_{\ell,-m} = (-1)^\ell h^*_{\ell,m} \) is broken \citep{Boyle:2014ioa}, meaning the negative \( m \)-modes are no longer directly related to the positive \( m \)-modes. Precession also causes modulations of  the amplitude and frequency of the plus and cross polarizations \citep{CalderonBustillo:2016rlt}, leading to an effective mode mixing \citep{Schmidt:2012rh,Ossokine:2013zga}. This allows for an even more accurate measurement of the inclination angle, helping to resolve the degeneracy with the luminosity distance, which in turn significantly improves distance and sky localization estimates when both HOMs and precessing spins are detected \citep{London:2017bcn,Chen:2018omi,Vitale:2018wlg}.

%
Including all spin parameters in our analysis, thereby allowing for spin precession, is crucial for better constraining source parameters and resolving degeneracies. However, spin precession introduces frequency-domain modulations, making the data more complex. To address this, we need to improve our method performance, which may require training {\sc Dingo-IS} with group-equivariant NPE \citep[GNPE;][]{Dax:2021myb} or flow-matching posterior estimation \citep[FMPE;][]{2023mla..confE..34W}.

GNPE exploits physical symmetries in GW signals, such as arrival time differences between detectors, by using an additional neural network to learn and apply time shifts to both parameters and data. This enables {\sc Dingo-IS} to focus on smaller time shifts, which correspond to minor phase shifts in the frequency domain, simplifying the data stream and improving inference. Flow matching, a recent generative modeling technique, enhances the flexibility of network architectures for handling complex data \citep{lipman2023flowmatchinggenerativemodeling}.

In this section, we investigate the impact of HOMs and precessing spins on our results using the standard inference code {\sc Bilby} with {\sc nessai} stochastic sampling algorithm. To study the impact of HOMs, we generate a new injection with the same parameters $\theta^{\mathrm{inj}}$ of \texttt{Event 1} (see Sec.~\ref{sec:single}), but without noise, meaning that in Eq.~\ref{eq:likelihood}, we set $d=h(\theta^{{\mathrm{inj}}})$. 

We generate waveforms using the {\sc IMRPhenomXPHM} approximant \citep{PhysRevD.103.104056}, including all available HOMs and, separately, only the dominant quadrupole mode $(\ell,m)=[(2,2),(2,-2)]$. We perform inference with the same sampling setup, using $5000$ live points on four CPUs. The sampling time differs significantly: about 10 hours for HOMs with $3.7 \times 10^7$ likelihood evaluations, while the dominant mode requires 13 days and 18 hours with $8.5 \times 10^8$ likelihood evaluations. This is because the quadrupole mode cannot resolve degeneracies, making sampling slower and PE less precise, as shown in Fig.~\ref{fig:hom}. These results align with previous findings for current gravitational wave interferometers \citep{LIGOScientific:2020stg} and LISA \citep{Marsat:2020rtl}, confirming the importance of HOMs for PE with ET \citep{Fairhurst:2023beb}.

\begin{figure}
    \centering
    \includegraphics[width=1\linewidth]{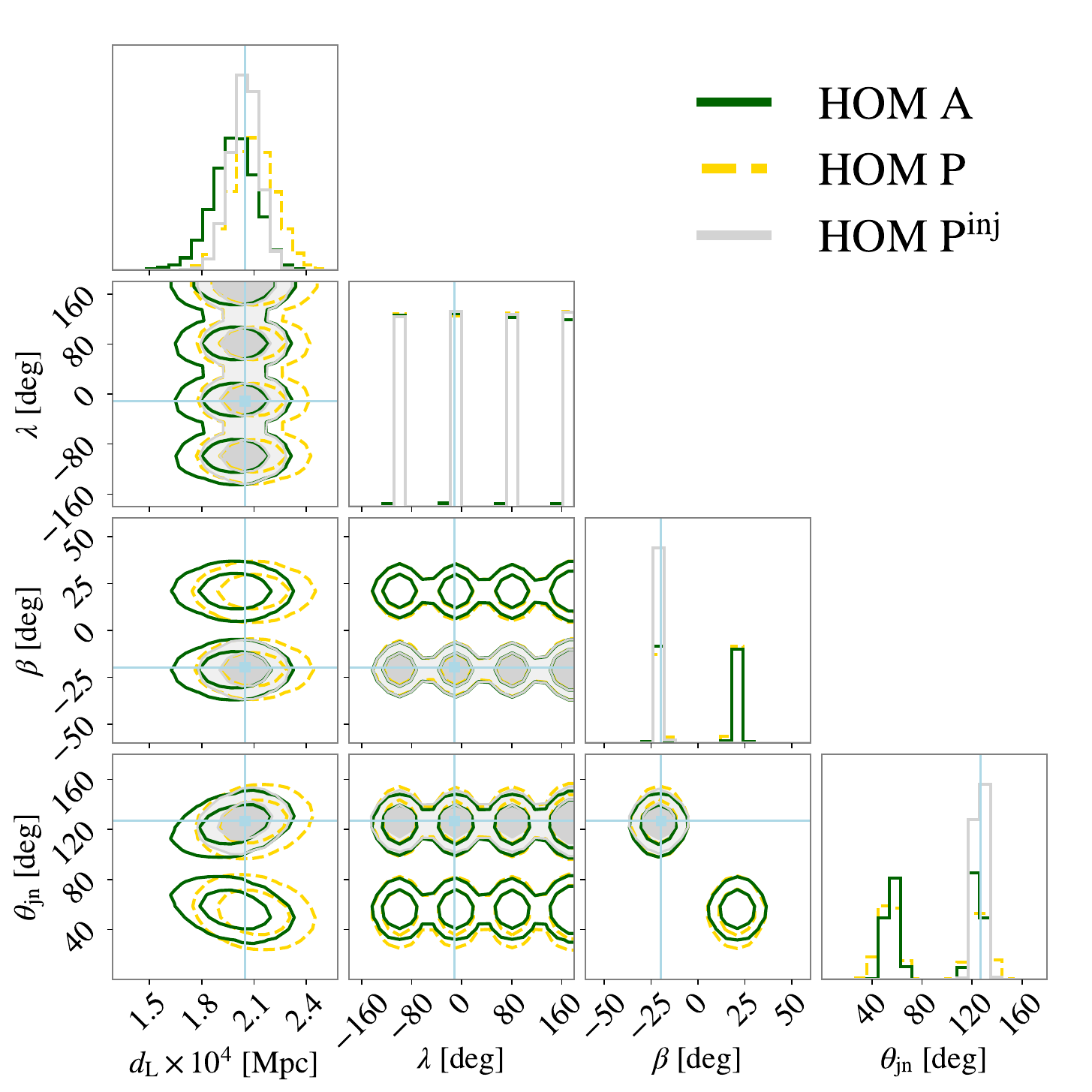}
    \caption{
    Marginalized one- and two-dimensional posterior distributions for luminosity distance ($d_{\mathrm{L}}$), azimuth ($\lambda$), altitude ($\beta$), and inclination angle ($\theta_{\mathrm{jn}}$) for \texttt{Event 1} with zero noise. Waveform generation is done using all higher-order modes (HOMs) available with {\sc IMRPhenomXPHM} approximant. The posterior distributions correspond to three cases: HOM A (dark green) with aligned spins, HOM P (yellow) where precessing spin parameters are sampled, and HOM P$^{\mathrm{inj}}$ (grey) where precessing spin parameters are fixed to their injected values. Contours indicate the 68\% and 95\% credible regions. Vertical and horizontal lines mark the true injected values. See Sec.~\ref{sec:hom} for details.}
    \label{fig:prec}
\end{figure}

We analyze a second injection, again without noise, using the same parameters of \texttt{Event 1} but modifying the spin angles. In this case, we introduce precessing spins with the following parameters: $\theta_1 = 0.5$ rad, $\theta_2 = 1.0$ rad, $\phi_{12} = 1.7$ rad, and $\phi_{jl} = 0.3$ rad. Here, $\phi_{12}$ is the angle between the projections of the two spin vectors in the orbital plane, while $\phi_{jl}$ is the angle between the orbital and total angular momentum (see Fig.~7 of \citealt{Dupletsa:2024gfl}). We consider two inference setups: in the first, we fix the spin angles to their injected values by using delta-function priors; in the second, we allow $\theta_1$, $\theta_2$, $\phi_{12}$, and $\phi_{jl}$ to be freely sampled. 

The outcome of this test is shown in Fig.~\ref{fig:prec}. When the precessing spin parameters are perfectly known, i.e., when the spin angles are fixed to their injected values, the degeneracy between the inclination angle and luminosity distance is resolved. As a consequence, the number of sky modes discussed in Sec.~\ref{sec:skyloc} is reduced to four. In particular, measuring sufficient spin precession removes the degeneracy between inclination angle ($\theta_{\mathrm{jn}}$) and altitude ($\beta$), which means the sky modes remain degenerate in the azimuth angle ($\lambda$). However, accurately constraining the precessing spin parameters is not guaranteed (see Fig.~\ref{fig:prec}), even for an high optimal SNR signal ($\rho \sim 83$) like the one considered in our case. This aligns with previous research showing that the ability to estimate precession depends heavily on the specific characteristics of the source \citep{Mills:2020thr,2024AAS...24345609R}, including signal duration, with longer signals providing better constraints \citep{Romero-Shaw:2022fbf}.


\subsection{Astrophysical population}
\label{sec:astro}

\begin{figure}
    \centering
    \includegraphics[width=1\linewidth]{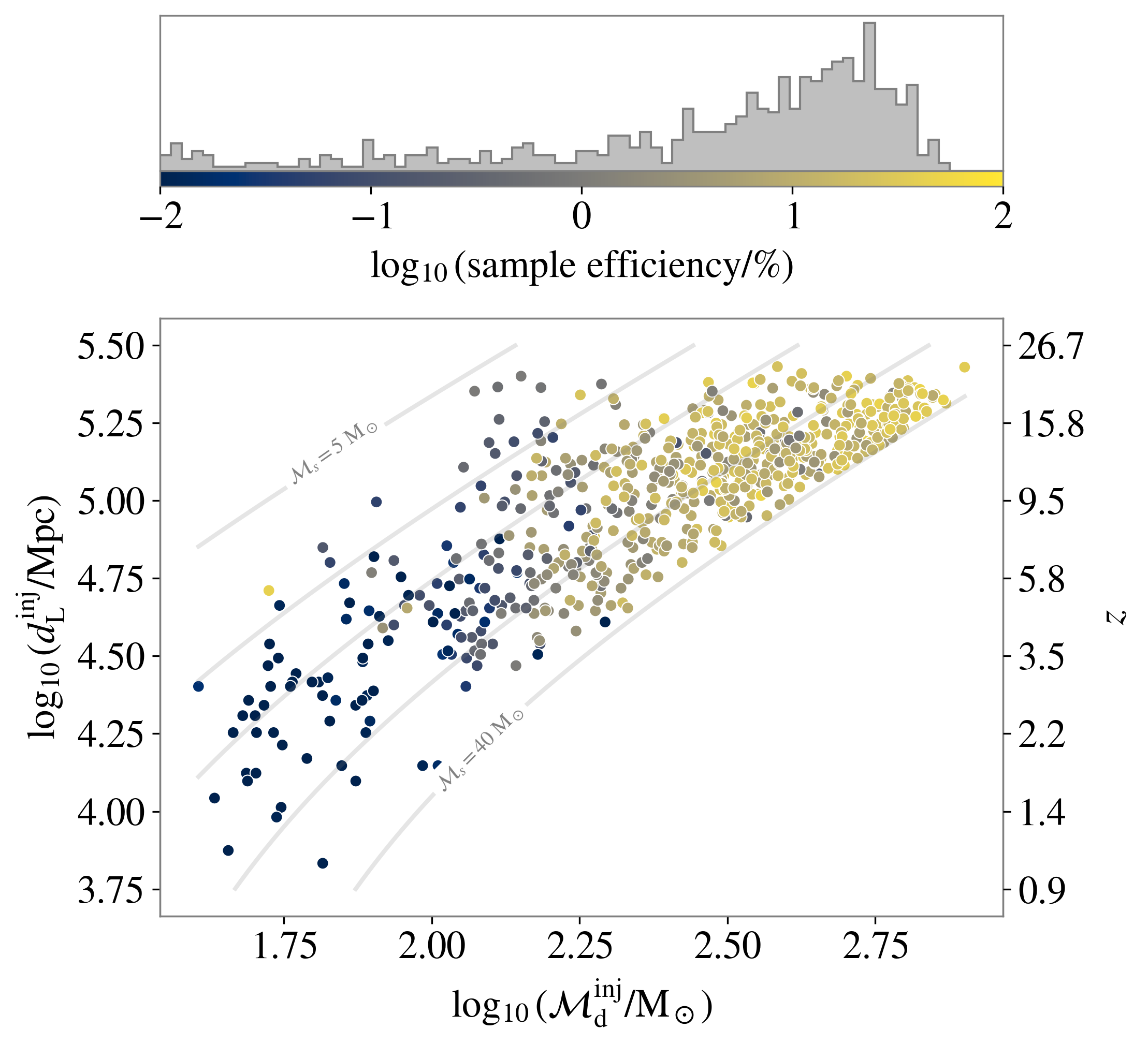}
    \caption{
    Injections from an astrophysical distribution of binary black holes (BBHs) formed from Population III stars, shown as a function of detector-frame chirp mass ($x$-axis) and luminosity distance (left $y$-axis), with corresponding redshift values (right $y$-axis). Each point is color-coded by the sample efficiency, and its marginal distribution is shown as a gray histogram above. The gray lines represent constant chirp masses in the source frame, with values of $\mathcal{M}_s = 5,\, 10,\, 15,\, 25,\, 40$ M$_\odot$. For more details, see Sec.~\ref{sec:astro}.}
    \label{fig:sample_efficiency_astro}
\end{figure}



We estimate the parameters of GW events extracted from an astrophysical population, which is compatible with the prior ranges listed in Table~\ref{tab:prior}. Specifically, we focus on BBHs formed from Pop.~III stars, which are believed to form and merge at high redshift \citep{Liu:2024mkh,Mestichelli:2024djn}. Our analysis includes one year of BBH mergers predicted by the \textsc{LOG1} model \citep{costa2023, santoliquido2023}. This model assumes a flat-in-log initial mass function \citep{Stacy:2012iz, Tanikawa:2020cca, Prole:2021nym} and adopts the mass ratio, eccentricity and orbital period of zero-age main sequence progenitor stars from \citet{Sana:2012px}. 

Single and binary stellar evolution processes are modeled using the population synthesis code \textsc{sevn} \citep{2023MNRAS.524..426I}. The merger rate density of Pop.~III BBHs is computed with \textsc{cosmo$\mathcal{R}$ate} \citep{Santoliquido:2020bry, Santoliquido:2020axb}, which combines population synthesis results with the star formation rate density model of Pop.~III stars from \citet{Hartwig:2022lon}. For further details, see \citet{Santoliquido:2024oqs}.

To create our injected BBHs from Pop.~III stars, we take only the chirp mass, mass ratio, and luminosity distance predicted with the \textsc{LOG1} model. The other waveform parameters are sampled from their prior distributions (Sec.~\ref{sec:prior}). The \textsc{LOG1} model predicts about 690 BBH mergers per year, regardless of whether they are detectable with ET-$\Delta$. We processed all of them with {\sc Dingo-IS}, except for 14 sources with an injected detector-frame chirp mass $\mathcal{M}^{\mathrm{inj}}_{\mathrm{d}} < 40$ M$_\odot$ and one source with an injected luminosity distance $d^{\mathrm{inj}}_\mathrm{L} < 5\times10^3$ Mpc. For each source, we extracted $2 \times 10^4$ samples. Figure~\ref{fig:sample_efficiency_astro} confirms the results shown in Fig.~\ref{fig:sample_efficiency}, where injections were sampled from the priors distribution. The average and median sample efficiency are 11.4 \% and 8.1\%, respectively, with $\sim 75$\% of sources with sample efficiency $> 1\%$. In particular, sample efficiency is higher for sources merging at high redshift, reaching an average sample efficiency of 16\% at $z > 10$.


Beyond BBHs formed from Pop.~III stars, intermediate-mass black holes (IMBHs) represent another class of objects with masses in the range specified in Table~\ref{tab:prior} \citep{2020ARA&A..58..257G, 2023arXiv231112118A}. In the $10^2$--$10^3$ M$_\odot$ range, IMBHs are thought to primarily form through stellar collisions \citep{2002ApJ...576..899P,2015MNRAS.454.3150G,2016MNRAS.459.3432M,2021A&A...652A..54A,2023MNRAS.526..429A}, hierarchical BH mergers \citep{2002MNRAS.330..232C,2004ApJ...616..221G,2019MNRAS.486.5008A,2021A&A...652A..54A}, close interactions between stars and BHs \citep{2017MNRAS.467.4180S,2023MNRAS.521.2930R,2023MNRAS.526..429A}, or a combination of these processes \citep{2015MNRAS.454.3150G,2023MNRAS.526..429A}. However, due to limited observational evidence, the mechanisms governing IMBH formation and growth remain unclear. In this context, ET and 3G detectors will provide crucial insights into the nature of IMBHs \citep{2020CQGra..37u5011A,2024arXiv240207571C,2025JCAP...01..108A,2025arXiv250312263A,Reali:2024hqf}.

\subsection{Implications}
\label{sec:impli}

The results shown in this work have important implications. For instance, although ET may observe BBH mergers up to $z \sim 100$ \citep{Ng:2020qpk,Ng:2021sqn,Kalogera:2021bya,Yi:2022tzs,Ng:2022agi,Branchesi:2023mws}, several studies have highlighted the difficulty ET will face in determining the luminosity distance distribution of high-redshift sources \citep{Iacovelli:2022bbs,Mancarella:2023ehn,Marcoccia:2023khb,Santoliquido:2024oqs}. However, these studies predominantly used the FIM approximation. We expect that their conclusions could be more optimistic if PE were done with NPE, as presented in this work (see Fig.~\ref{fig:scatter}).

NPE demonstrates the realistic sky localization performance of the triangular ET for short-lived high-redshift sources, showing that for well-localized sources, where $\Delta \Omega_{90\%} < 100$ deg$^2$, the uncertainties are up to 8 times larger than those predicted using the FIM (see Fig.~\ref{fig:scatter}). This might have significant implications for dark siren cosmology, which involves statistically inferring the redshift from potential host galaxies within the estimated GW localization volume \citep{Schutz:1986gp,DelPozzo:2011vcw,Chen:2017rfc,Libanore:2020fim,Gair:2022zsa,Bosi:2023amu,Borghi:2023opd}. Several studies have predicted the ability of ET to constrain at percent level cosmological parameters such as $H_0$ and $\Omega_m$ \citep{Muttoni:2021veo,Muttoni:2023prw,Chen:2024gdn,Califano:2025qbx} primarily relying on sources at $z < 1$, where galaxy catalogs are expected to be complete. These studies used the FIM to estimate the uncertainty in sky localization. Their results could be more optimistic than they would be with NPE, unless the sky localization is evaluated considering a network of 3G detectors. 

Sky modes will only be seen for short-lived sources, where antenna patterns are time-independent. For low-mass sources, such as binary neutron stars (BNSs), which could last hours in the ET frequency band, we expect sky modes will be resolved once Earth's rotation is taken into account during parameter estimation \citep{Chan:2018csa,Nitz:2021pbr,Li:2021mbo,Li:2024ruk,Singh:2021bwn}.




\subsection{Energy cost}
\label{sec:energy}

Any improvement in PE for 3G detectors, such as the NPE method presented here, must be both time-efficient to meet the scientific requirements of ET and energy-efficient to ensure sustainability \citep{2024PhRvD.110h3033W}. This is important not only for making scientific research more environmentally friendly but also for understanding the energy demands of PE in preparation for future ET facilities, including power supply needs.

We estimate the energy cost in kilowatt-hours (kWh) for training {\sc Dingo-IS} and performing PE on $10^3$ injections, comparing it to the estimated energy cost of {\sc Bilby}. 
The analysis is run on an NVIDIA A100 80 GB GPU, with a maximum power consumption of 
300 W, and on multiple AMD EPYC 7513 CPUs, with 32 cores each and a maximum power consumption of 
200 W.

Training takes six days using one GPU and 32 CPUs, consuming 970 kWh. Each injection is processed with eight CPUs for about 1.7 minutes on average for a total of $\sim 29$ hours, consuming up to 45 kWh. In total, training and inference require 1015 kWh. For comparison, an injection processed with {\sc Bilby} running on four CPUs, with \texttt{nlive=2000}, and using the {\sc nessai} sampler takes about five hours on average, resulting in an estimated total energy consumption of 4000 kWh for all processed injections.

This simple estimate highlights that NPE is not only faster, but can be also significantly more energy-efficient than traditional methods.



\subsection{Caveats}
\label{sec:caveats}

A major drawback of likelihood-free inference methods like {\sc Dingo-IS} is that if the training settings need to be changed, such as switching the waveform approximant, a new neural network must be trained using simulations generated with the updated settings. 
Training is not only the most time-consuming but also the most energy-intensive process of using NPE, as discussed in Sec.~\ref{sec:energy}.

To accelerate training and improve convergence of {\sc Dingo-IS}, we made a few simplifying assumptions. We set a low-frequency limit of $f_{\mathrm{min}} = 6$ Hz, even though ET is expected to be sensitive down to $\sim 2$ Hz. Reducing $f_{\mathrm{min}}$ allows for processing longer signals—up to 64 s for mergers happening at $f < f_{\mathrm{max}} = 256$ Hz—but requires a finer frequency resolution of $\mathrm{d}f = 1/64$ Hz. This increases the input array size by a factor of 8, making training more challenging. Methods for processing longer signals with {\sc Dingo-IS} have been developed in \cite{Dax:2024mcn}, where they include the use of multibanding and heterodyning to handle BNS signals up to one hour in length.

Additionally, we did not consider contamination from overlapping signals or transient non-Gaussian noise, which could complicate inference \citep{Raymond:2024xzj}. Addressing these limitations in future work will help establish NPE as a robust method for processing the full range of detectable sources with ET.

A comprehensive comparison of full PE performance across different ET designs is planned for future work, including an investigation of the potential benefits of the 2L configuration, where the increased baseline between detectors may help reduce multimodalities in sky localization \citep{Fairhurst:2009tc,Fairhurst:2017mvj,Grover:2013sha,Berry:2014jja}.

\section{Conclusions}
\label{sec:conclusions}

The Einstein Telescope, along with other third-generation detectors such as Cosmic Explorer \citep{Reitze:2019iox,Evans:2023euw,Gupta:2023lga} and next-generation detectors like LISA \citep{LISA:2024hlh}, will define the future of gravitational wave astronomy in the coming decades \citep{Punturo:2010zz,ET:2019dnz,Branchesi:2023mws}. These detectors are expected to observe thousands of sources every day \citep{2024PhRvD.110h3040B}. However, this enormous volume of data will pose significant challenges for data analysis and parameter estimation \citep[PE;][]{Couvares:2021ajn}. Current inference algorithms, which rely on stochastic sampling \citep{Ashton:2018jfp}, will struggle to handle this scale of data efficiently and rapidly, even with techniques that implement likelihood simplification \citep{2024PhRvD.110h4085N,2024arXiv241202651H} or advanced sampling methods \citep{2023MLS&T...4c5011W,2023ApJ...958..129W}. Therefore, the era of 3G detectors requires new technologies for PE, which must be developed now to guide the design of these detectors. 

To tackle this challenge, we focus on providing fast and accurate PE for a specific category of sources which is pivotal in the science case of 3G detectors: high-redshift BBH mergers \citep{Franciolini:2022tfm,Ng:2020qpk,Ng:2021sqn,Ng:2022agi,santoliquido2023,Santoliquido:2024oqs,Reali:2024hqf}. These short-lived sources are well-suited for being processed with Neural Posterior Estimation \citep[NPE;][]{Papamakarios:2016ctj}, a likelihood-free inference method implemented in {\sc Dingo-IS} \citep{Green:2020dnx,Dax:2021myb,Dax:2024mcn}. NPE learns the posterior probability distribution directly from simulated data using normalizing flows \citep{Kobyzev_2021}. 

We demonstrate in this work that NPE is a promising approach for 3G detector applications. With sufficient training and samples refined through importance sampling (see Sec.~\ref{sec:dingo}), the posterior distributions obtained from NPE are statistically indistinguishable from those derived using standard inference methods (Sec.~\ref{sec:single} and Fig.~\ref{fig:bilby}), at a fraction of time and energy cost (Sec.~\ref{sec:energy}).

The triangular configuration of ET (ET-$\Delta$, Fig.~\ref{fig:ETmap}) adopted in this work reveals unique characteristics in the posterior probabilities of detected sources, particularly in sky localization. We show that ET-$\Delta$ will observe up to eight degenerate sky modes for short-lived distant sources (Sec.~\ref{sec:skyloc} and Fig.~\ref{fig:horizon}), which can be explained by time- and frequency-independent antenna patterns. 

We analyze $1000$ simulated BBH signals extracted from the prior distribution (Sec.~\ref{sec:pop}) and achieve high accuracy in parameter estimation across a wide redshift range ($1 \lesssim z \lesssim 45$). The sampling efficiency is generally higher for more distant BBHs, averaging 12.5\% overall and increasing to about 20\% for sources merging at $z > 10$ (Fig.~\ref{fig:sample_efficiency}). This trend is confirmed in an analysis of 690 sources from a one-year observation of BBHs formed from Population~III stars \citep{costa2023,santoliquido2023,Santoliquido:2024oqs}, where the average sampling efficiency for $z > 10$ is about 16\% (Sec.~\ref{sec:astro} and Fig.~\ref{fig:sample_efficiency_astro}).

This study compares parameter estimation accuracy obtained with NPE and with the FIM formalism for the triangular ET configuration (Sec.~\ref{sec:pop}). 
FIM tends to overestimate luminosity distance errors for sources merging at $z > 10$, with discrepancies reaching up to a factor of 4. In sky localization, FIM fails to capture the eight degenerate sky modes characteristic of a triangular ET, significantly underestimating errors. For other parameters, such as the detector-frame chirp mass, mass ratio, and aligned spins, the FIM generally underestimates the uncertainties. These results pertain specifically to the science case investigated in this work, namely high-redshift BBH mergers.

These findings highlight the limitations of FIM in accurately capturing complex parameter dependencies as they can emerge in observations with a single ET triangle. However, as discussed in Sec.~\ref{sec:impli}, when time- and frequency-dependent antenna patterns are taken into account \citep{Singh:2021bwn,Piro:2022zos}, FIM analysis is expected to provide sky localizations comparable to those obtained with full PE \citep{Chan:2018csa,Nitz:2021pbr,Li:2021mbo,Li:2024ruk,Singh:2021bwn}. 


\section*{Data Availability}

The main data used in this study are publicly available on Zenodo \citep{santoliquido_2025_17358571} (\href{https://zenodo.org/records/17358571}{https://zenodo.org/records/17358571}). The latest public release of {\sc Dingo}  can be accessed from \href{https://github.com/dingo-gw/dingo}{https://github.com/dingo-gw/dingo}, while {\sc GWFish+Priors} is also openly available at \href{https://github.com/janosch314/GWFish}{https://github.com/janosch314/GWFish}. Additional data and code are available from the corresponding authors upon reasonable request.


\acknowledgments
We thank the anonymous referee for their careful reading of our manuscript and for the valuable suggestions that helped improve it. We thank Michael Williams, Luca Reali, Francesco Iacovelli, Andrea Maselli, Lavinia Paiella, Cristiano Ugolini, Benedetta Mestichelli, Alessio Ludovico De Santis, and Matteo Schulz for insightful discussions.
F.S. acknowledges financial support from the AHEAD2020 project (grant agreement no. 871158). M.B. acknowledges financial support from the Italian Ministry of University and Research (MUR) for PRIN grant METE under contract no. 2020KB33TP. MAS acknowledges funding from the European Union’s Horizon 2020 research and innovation programme under the Marie Skłodowska-Curie grant agreement no.~101025436 (project GRACE-BH, PI: Manuel Arca Sedda). MAS acknowledge financial support from the MERAC foundation. S.R.G. was supported by a UK Research and Innovation (UKRI) Future Leaders Fellowship (grant no.~MR/Y018060/1). I.M.R.-S. acknowledges the support of the Herchel Smith fund. E.B. is supported by NSF grants no. AST-2307146, no. PHY-2207502, no. PHY-090003 and no. PHY-20043; by NASA grant no. 21-ATP21-0010; by the John Templeton Foundation grant no. 62840; by the Simons Foundation; and by the Italian Ministry of Foreign Affairs and International Cooperation grant no. PGR01167. The research leading to these results has been conceived and developed within the Einstein Telescope Observational Science Board ET-0130A-25.

\appendix





\section{Priors}
\label{app:priors}

\begin{figure*}
    \centering
    \includegraphics[width=0.75\linewidth]{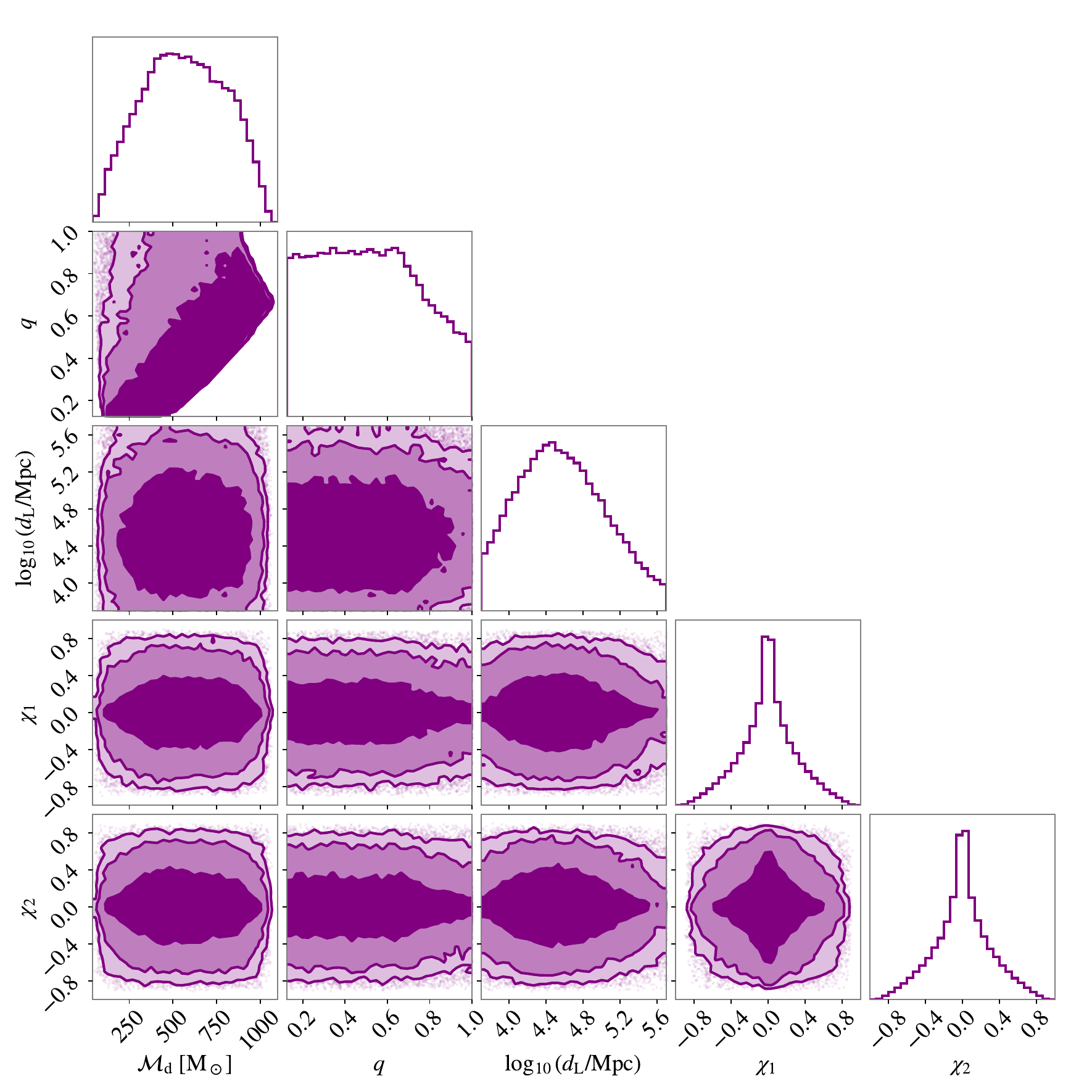}
    \caption{Marginalized one- and two-dimensional distributions of the injection parameters, sampled according to the chosen priors listed in Table~\ref{tab:prior}. Contours indicate the 68\%, 95\%, and 99\% credible regions. For details, see Sec.~\ref{sec:prior} and Appendix~\ref{app:priors}.}
    \label{fig:priors}
\end{figure*}

Fig.~\ref{fig:priors} shows the distribution of injected parameters sampled according to the chosen priors as presented in Sec.~\ref{sec:prior} and listed in Table~\ref{tab:prior}. We show only the nontrivial parameters like detector-frame chirp mass, mass ratio, luminosity distance, and aligned spins, since the other parameters are distributed uniformly in the specified ranges. The primary (secondary) detector-frame mass is constrained to be within 50 M$_\odot$ and 1500 M$_\odot$ (1000 M$_\odot$). This explains the sharp edges in the $p(\mathcal{M}_d,q)$ distribution in Fig.~\ref{fig:priors}.

\section{Event 2}
\label{app:event2}

\begin{figure*}
    \centering
    \includegraphics[width=1\linewidth]{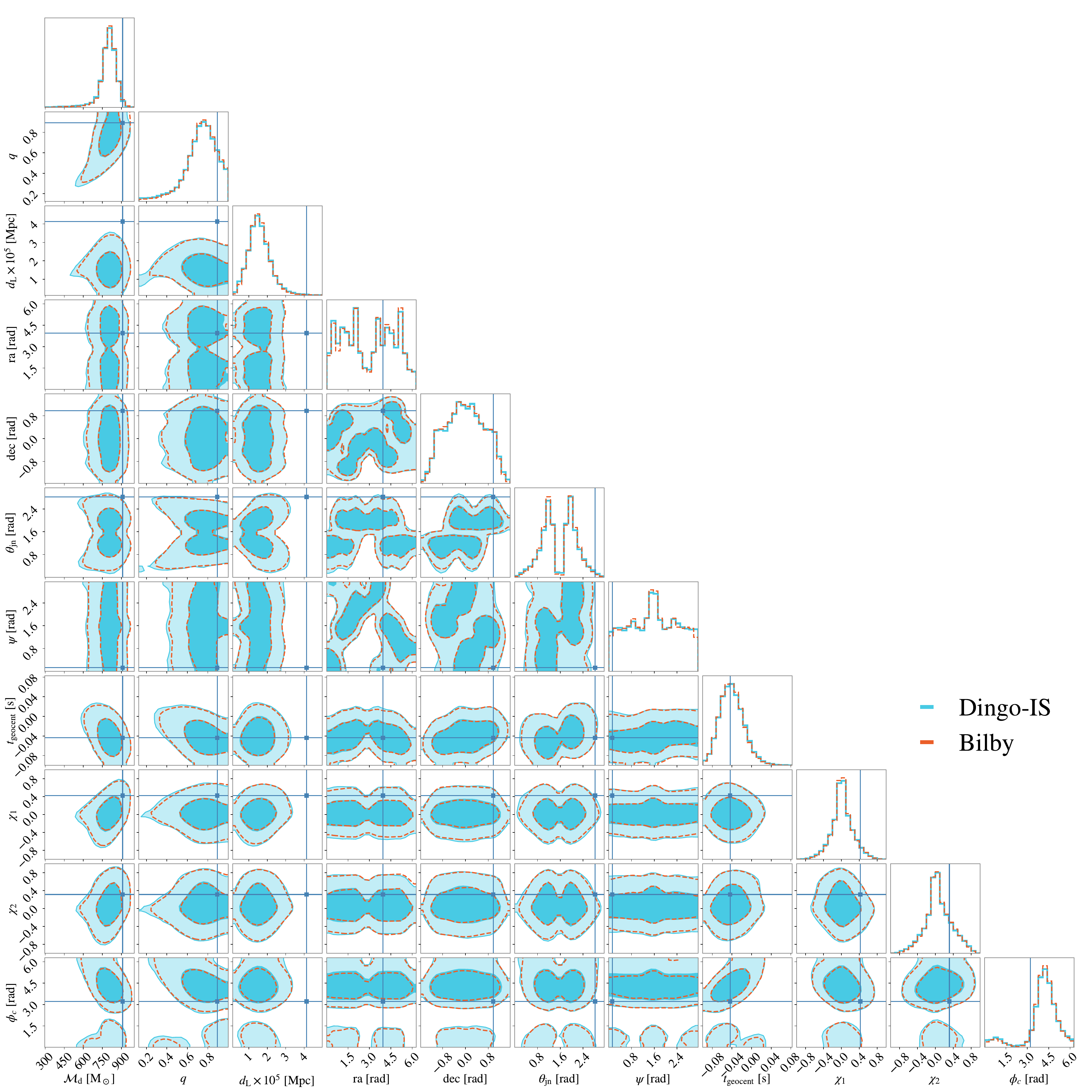}
    \caption{Marginalized one- and two-dimensional posterior distributions for \texttt{Event 2} over all set of parameters, comparing {\sc Dingo-IS} (light blue) and {\sc Bilby} (orange). Vertical and horizontal lines mark the true injected values (see Table~\ref{tab:event2}). Contours represent 68\% and 95\% credible regions. Sky modes are discussed in Sec.~\ref{sec:skyloc}. See Appendix~\ref{app:event2} for other details.}
    \label{fig:event2}
\end{figure*}

For completeness, we present the same analysis as in Sec.~\ref{sec:single} but for a different injection, \texttt{Event 2}, which has a higher injected redshift and therefore a lower SNR with respect to \texttt{Event 1}. We draw $1 \times 10^5$ samples from the proposal distribution $q(\theta|d)$. The sampling efficiency for \texttt{Event 2} is approximately $25\%$, corresponding to $\sim 2.5 \times 10^4$ effective samples. 

In {\sc Bilby}, we set the number of live points to \texttt{nlive = 10\_000}. 
The default settings in {\sc{nessai}} do not yield satisfactory results for this source. To improve robustness, we increase the {\texttt{volume\_fraction}} from 0.95 to 0.98. This parameter determines the fraction of the total probability enclosed by the contour in latent space—the base space from which samples are initially drawn before being mapped to physical space via the normalizing flow \citep{2021PhRvD.103j3006W}. While a higher \texttt{volume\_fraction} enhances robustness, it also increases computational cost. With these settings, {\sc{Bilby}} requires approximately 1 day and 20 hours to reach convergence.

We follow the same procedure outlined in Sec.~\ref{sec:single} to define a threshold on the JSD for assessing statistical significance. In this case, we generate with {\sc{Dingo-IS}} 100 sets of $10^5$ posterior samples each.

As shown in Table~\ref{tab:event2}, the median JSD values for the one-dimensional marginal distribution of all parameters 
are 
below the corresponding JSD$^{\mathrm{thr}}$, indicating that the posteriors obtained with {\sc Dingo-IS} and {\sc Bilby} are identical. 
This agreement is also evident in Fig.~\ref{fig:event2}, which displays the one- and two-dimensional marginal posteriors. 

The log-evidence estimates are $\log \mathcal{Z}(d) = -5917.651 \pm 0.005$ for {\sc Dingo-IS} and $-5918.13 \pm 0.04$ for {\sc Bilby}
, a 
difference within statistical uncertainty. 


\begin{table}[h]

    
    \centering
    \begin{tabular}{llccl}
    \toprule
        Parameter & $\theta^{\mathrm{inj}}$ &  $\langle\mathrm{JSD}\rangle$ & JSD$^{\mathrm{thr}}$ & $\theta$  \\
        \hline
         &  &  &  \\
        $\mathcal{M}_{\mathrm{d}}$ & 909 M$_\odot$ & 
        1.3 & 2.3 & ${805}_{-103}^{+80}$  M$_\odot$\\
        $q$ & 0.89 & 
        1.1 & 2.3 & ${0.8}_{-0.3}^{+0.2}$ \\
        $d_{\mathrm{L}}$ & 413 Gpc & 
        1.6 & 2.6 & ${153}_{-83}^{+108}$ Gpc\\
        ra & 3.94 rad & 
        1.0 & 2.7 & ${3.20}_{-2.82}^{+2.35}$ rad\\
        dec &  0.97 rad & 
        1.5 & 3.0 & ${-0.00}_{-1.03}^{+1.04}$ rad\\
        $\theta_{\mathrm{jn}}$ & 2.82 rad & 
        2.2 & 2.7 & ${1.73}_{-0.97}^{+0.72}$ rad\\
        $\phi_c$ & 3.21 rad & 
        1.6 & 2.9 & ${4.22}_{-3.17}^{+1.22}$ rad\\
        $\psi$ & 0.13 rad & 
        1.2 & 3.0 & ${1.60}_{-1.36}^{+1.35}$ rad \\
        $t_{\mathrm{geocent}}$ & -0.04 s & 
        1.0 & 2.8 &  ${0.04}_{-0.03}^{+0.04}$ s\\
        $\chi_1$ & 0.42 & 
        1.3 & 2.8 & ${0.01}_{-0.34}^{+0.41}$\\   
        $\chi_2$ & 0.31 & 
        1.1 & 2.5 & ${0.04}_{-0.43}^{+0.54}$\\   
         &  &  &\\   
        $m_{1,\mathrm{s}}$ & 31 M$_\odot$ & & & $72_{-26}^{+60}$ M$_\odot$\\
        $m_{2,\mathrm{s}}$ & 28 M$_\odot$ & & & ${53}_{-25}^{+53}$ M$_\odot$\\
        $z$ & 34 & & & $14.0_{-7.0}^{+8.6}$ \\
        $\rho$ & 14 &\\
        $\Delta \Omega_{90\%}$ & & & & 22 922 deg$^2$  \\
    \bottomrule
    \end{tabular}
    \caption{Injected parameters $\theta^{\mathrm{inj}}$ for \texttt{Event 2}, along with the median Jensen-Shannon divergence ($\langle\mathrm{JSD}\rangle$, in units of $10^{-3}$ nat), which quantifies the deviation between {\sc Dingo-IS} and {\sc Bilby} for one-dimensional marginal posteriors. For comparison, the corresponding JSD threshold (JSD$^{\mathrm{thr}}$) is also shown. The last column on the right shows the median and 90\% credible interval of the recovered parameters $\theta$ with {\sc Dingo-IS}. We report in the bottom rows the  source-frame primary and secondary mass, redshift, the optimal signal-to-noise ratio ($\rho$, see Eq.~\ref{eq:optimalSNR}), and the sky localization error. See Appendix~\ref{app:event2} for details.}
    \label{tab:event2}
\end{table}

\bibliographystyle{apsrev4-2}
\bibliography{main}

\end{document}